\documentclass[aps,prc,preprint,superscriptaddress,preprintnumbers]{revtex4-1}

\usepackage{textcomp}
\usepackage{bm}
\usepackage{amsmath}
\usepackage{mathrsfs}
\usepackage{graphicx}
\usepackage{xcolor}
\usepackage{setspace}
\usepackage{booktabs}
\usepackage{multirow}

\usepackage{CJKutf8}

\newcommand{\te}{\mathrm{e}}
\newcommand{\ti}{\mathrm{i}}

\begin{document}
\begin{CJK}{UTF8}{gbsn}

\title{Nuclear matrix element of neutrinoless double-$\beta$ decay: \\ Relativity and short-range correlations}

\author{L. S. Song (宋凌霜)}
\affiliation{State Key Laboratory of Nuclear Physics and Technology, School of Physics, Peking University, Beijing 100871, China}
\author{J. M. Yao (尧江明)}
\affiliation{Department of Physics and Astronomy, University of North Carolina, Chapel Hill, NC 27516-3255, USA}
\affiliation{School of Physical Science and Technology, Southwest University, Chongqing 400715, China}
\author{P. Ring}
\affiliation{Physik-Department der Technischen Universit{\"a}t M{\"u}nchen, D-85748 Garching, Germany}
\affiliation{State Key Laboratory of Nuclear Physics and Technology, School of Physics, Peking University, Beijing 100871, China}
\author{J. Meng (孟杰)}
\affiliation{State Key Laboratory of Nuclear Physics and Technology, School of Physics, Peking University, Beijing 100871, China}
\affiliation{School of Physics and Nuclear Energy Engineering, Beihang University, Beijing 100191, China}
\affiliation{Department of Physics, University of Stellenbosch, Stellenbosch 7602, South Africa}


\begin{abstract}
\begin{description}
	\item[Background]
	The discovery of neutrinoless double-beta ($0\nu\beta\beta$) decay would demonstrate the nature of neutrinos, have profound implications for our understanding of matter-antimatter mystery, and solve the mass hierarchy problem of neutrinos. The calculations for the nuclear matrix elements $M^{0\nu}$ of $0\nu\beta\beta$ decay are crucial for the interpretation of this process.
	\item[Purpose]
	We study the effects of relativity and nucleon-nucleon short-range correlations on the nuclear matrix elements $M^{0\nu}$ by assuming the mechanism of exchanging light or heavy neutrinos for the $0\nu\beta\beta$ decay.
	\item[Methods]
	The nuclear matrix elements $M^{0\nu}$ are calculated within the framework of covariant density functional theory, where the beyond-mean-field correlations are included in the nuclear wave functions by configuration mixing of both angular-momentum and particle-number projected quadrupole deformed mean-field states.
	\item[Results]
	The nuclear matrix elements $M^{0\nu}$ are obtained for ten $0\nu\beta\beta$-decay candidate nuclei.
	The impact of relativity is illustrated by adopting relativistic or nonrelativistic decay operators.
	The effects of short-range correlations are evaluated.
	\item[Conclusions]
    The effects of relativity and short-range correlations play an important role in the mechanism of exchanging heavy neutrinos though the influences are marginal for light neutrinos.
	Combining the nuclear matrix elements $M^{0\nu}$ with the observed lower limits on the $0\nu\beta\beta$-decay half-lives, the predicted strongest limits on the effective masses are $|\langle m_\nu\rangle|<0.06~\mathrm{eV}$ for light neutrinos and $|\langle m_{\nu_h}^{-1}\rangle|^{-1}>3.065\times 10^8~\mathrm{GeV}$ for heavy neutrinos.
\end{description}
\end{abstract}

\maketitle


\section{Introduction}\label{intro}

The neutrinoless double-$\beta$ ($0\nu\beta\beta$) decay is a process where an even-even nucleus $(N, Z)$ transforms into its even-even neighbor $(N-2, Z+2)$ with only two electrons emitted.
The fact that the $0\nu\beta\beta$ decay violates the total lepton number by two units makes it a probe sensitive to revealing the mysterious nature of massive neutrinos: This process occurs only if the neutrinos are Majorana particles and the violation of total lepton number is possible.
Several other fundamental questions on neutrinos, including their absolute mass scale, mass spectrum hierarchy (normal, inverted, or quasidegenerate), and the mechanism of masses generation, are expected to be clarified if one can possibly combine the results from this process and other neutrino experiments~\cite{Bilenky2010}.
To date, no actual signal for the $0\nu\beta\beta$ decay has been confirmed despite numerous experimental data released.
Recently, the most stringent lower limits on the half-lives have been reported by the KamLAND-Zen Collaboration~\cite{Gando2016} for ${}^{136}$Xe, $T_{1/2}^{0\nu}>1.07\times 10^{26}$ yr (90\% C.L.), and by the NEMO-3 Collaboration~\cite{Arnold2016} for ${}^{150}$Nd, $T_{1/2}^{0\nu}>2.0\times 10^{22}$ yr (90\% C.L.).

In the $0\nu\beta\beta$-decay mechanism of exchanging virtual Majorana neutrinos, the half-life $T_{1/2}^{0\nu}$ is inversely proportional to an effective parameter $f(m_i,U_{ei})$ related to neutrino masses, a kinematic phase-space factor $G_{0\nu}$, and the nuclear matrix element (NME) $M^{0\nu}$ squared:
\begin{eqnarray}
\label{eq:half-life}
	[ T_{1/2}^{0\nu}]^{-1}=G_{0\nu}\; g_A^4\;| M^{0\nu}|^2\;f(m_i,U_{ei}).
\end{eqnarray}
Considering the two limiting cases of neutrino propagator,
\begin{eqnarray}
\label{eq:propagator}
	\frac{m_i}{q_\mu q^\mu-m_i^2}\to
	\begin{cases}
		m_i/q_\mu q^\mu, & m_i^2\ll q_\mu q^\mu \\
		-1/m_i,          & m_i^2\gg q_\mu q^\mu
	\end{cases}
\end{eqnarray}
the amplitude is proportional to the mass for a light neutrino,
\begin{eqnarray}
\label{eq:flight}
	f(m_i,U_{ei})&=&|\langle m_\nu \rangle|^2 m_e^{-2},\\
	\langle m_\nu \rangle &=&\sum_k\left( U_{ek}\right)^2m_k,\notag
\end{eqnarray}
but inversely proportional to the mass for a heavy neutrino,
\begin{eqnarray}
\label{eq:fheavy}
	f(m_i,U_{ei})&=&|\langle m_{\nu_h}^{-1}\rangle|^2 m_p^2,\\
	\langle m_{\nu_h}^{-1}\rangle &=&\sum_{k_h}\left( U_{ek_h}\right)^2m_{k_h}^{-1}.\notag
\end{eqnarray}
Note that $q_\mu$ is the momentum transferred by the neutrino and $U_{ek}$ and $U_{ek_h}$ are elements in the neutrino mixing matrix that mix light and heavy neutrinos, respectively.
$m_e$ and $m_p$ are electron and nucleon masses, and the bare value $g_A= 1.254$ is used for the axial-vector coupling constant.
Given that the phase-space factor $G_{0\nu}$ has been precisely determined~\cite{Kotila2012}, an accurate knowledge of the NME $M^{0\nu}$ is the key to connecting the experimental measurement with fundamental physics.

The calculation of the NME requires the wave functions of initial and final nuclear states as well as the decay operator.
Previously, the NMEs $M^{0\nu}$ have been calculated within the framework of covariant density functional theory (CDFT)~\cite{Song2014, Yao2015, Ring2015, Yao2016, Meng2016}, where the relativistic wave functions and the relativistic $0\nu\beta\beta$-decay operator derived from weak interaction Hamiltonian are used in the calculations.
Various nonrelativistic nuclear structure models have been applied as well.
They include the configuration-interacting shell model (CISM)~\cite{Haxton1984, Wu1985, Retamosa1995, Caurier1996, Caurier2008, Menendez2009, Menendez2009a, Menendez2011, Senkov2013, Senkov2014, Senkov2014a, Horoi2010, Horoi2013, Neacsu2015, Horoi2016, Iwata2016}, the quasiparticle random phase approximation (QRPA)~\cite{Tomoda1987, Muto1989, Staudt1990, Stout1992, Staudt1992, Pantis1992, Kortelainen2007, Kortelainen2007a, Pantis1996, Simkovic1997, Simkovic1999, Rodin2006, *Rodin2007, Simkovic2008, Simkovic2009, Fang2010, Fang2011, Mustonen2013, Terasaki2015, Simkovic2013, Fang2015}, the projected Hartree-Fock-Bogoliubov (PHFB) model~\cite{Chaturvedi2008, Chandra2009, Rath2009, Rath2010, Rath2012, Rath2013}, the interacting boson model (IBM)~\cite{Barea2009, Barea2012, Barea2013,Barea2015}, and the nonrelativistic energy density functional (EDF) theory~\cite{Rodriguez2010, Rodriguez2011, Vaquero2013}.
In contrast with the CDFT application, the $0\nu\beta\beta$-decay operator has to be reduced to its nonrelativistic form in these calculations to be adapted to the nonrelativistic nuclear wave functions.
Therefore, the fully relativistic framework of CDFT allows one to examine the validity of the nonrelativistic approximation and to reveal the relativistic effects in the NME by conducting comparative studies with the relativistic or nonrelativistic-reduced decay operators, respectively.

Previous studies based on beyond-mean-field CDFT~\cite{Song2014, Yao2015} have shown that the nonrelativistic decay operator is a good approximation to the full relativistic operator within the assumption of light-neutrino exchange.
The goal of this paper is to generalize the calculations to the case with heavy-neutrino exchange and to present a comprehensive study on the effects of relativity and nucleon-nucleon short-range correlations (SRCs) on the NME of $0\nu\beta\beta$ decay.
The calculations are based on nuclear wave functions in which the dynamic effects of particle-number and angular-momentum conservations as well as shape fluctuations are incorporated by the projection techniques and the generator coordinate method (GCM), in full analogy to Refs.~\cite{Song2014, Yao2015}.
The SRC corrections neglected in previous calculations of light-neutrino NME are now taken into account via a Jastrow function using the Argonne V18 parametrization~\cite{Simkovic2009a, Jastrow1955, Miller1976}.

\section{Formalism}\label{theory}

In the framework of beyond-mean-field CDFT, the nuclear many-body wave function is constructed by superposing a set of quantum-number projected nonorthogonal states around the equilibrium shape~\cite{Yao2008,Yao2009,Yao2010,Yao2014,Yao2015a},
\begin{eqnarray}
\label{eq:GCMwf}
	\left|JNZ;\alpha \right\rangle=\sum_{\kappa\in\{{\beta_2,K}\}}f_\kappa^{J\alpha}\,\hat P_{MK}^J \hat P^N \hat P^Z \, |\beta_2\rangle.
\end{eqnarray}
The deformation parameters $\beta_2$ are chosen as the generator coordinates in the GCM method so that the quadrupole axial deformation and its quantum fluctuations are considered.
The reference states $|\beta_2\rangle$ are a set of BCS states generated from the self-consistent mean-field calculations based on the universal relativistic energy functional PC-PK1~\cite{Zhao2010}.
The projection operators $\hat P^G$'s ($G\equiv J,N,Z$)~\cite{Ring1980} are responsible for restoring broken symmetries by projecting the reference wave functions onto states with good angular momenta $J$ and numbers ($N, Z$) of neutrons and protons.
The coefficients $f_\kappa^{J\alpha}$ are determined by solving the Hill-Wheeler-Griffin equation~\cite{Ring1980}.
The indices $\alpha=1,2,\dots$ distinguish different nuclear states with energy $E_\alpha$.

The $0\nu\beta\beta$-decay operator is derived from the second-order weak Hamiltonian with charge-exchange nucleonic and leptonic currents. It reads
\begin{eqnarray}
\label{eq:operator}
	\hat{\mathcal O}^{0\nu}&=&\frac{4\pi R}{g_A^2} \iint d^3 x_1 d^3 x_2 \int\frac{d^3q}{(2\pi)^3}\, h(q)\notag\\
	&\times & \mathcal J_\mu^\dagger(\bm x_1)\mathcal J^{\mu\dagger}(\bm x_2)\;\te^{\ti\bm q\cdot(\bm x_1-\bm x_2)},
\end{eqnarray}
with $R=1.2A^{1/3}$ fm.

The neutrino potential $h(q)$ for light-neutrino exchange is
\begin{eqnarray}
\label{eq:hofq-light}
	h(q)&=&q^{-1}(q+E_d)^{-1}\,,\\
	E_d&\equiv & \bar E-(E_I+E_F)/2\,,\notag
\end{eqnarray}
where $E_{I(F)}$ corresponds to the energy of initial (final) nuclear state, and $\bar E$ is the average energy of intermediate states.
For heavy-neutrino exchange the neutrino potential is
\begin{eqnarray}
\label{eq:hofq-heavy}
	h(q)=(m_pm_e)^{-1}.
\end{eqnarray}
These potentials are obtained by taking the limiting forms of the neutrino propagator in Eq.~(\ref{eq:propagator}).
While the light-mass limit leads to a $q^{-2}$ dependence in $h(q)$, the heavy-mass limit gives a constant.

The charge-exchange nucleonic current is given by $\mathcal J_\mu^\dagger(\bm x)\equiv\bar\psi(\bm x)\Gamma_\mu(q)\tau_-\psi(\bm x)$, with the vertex,
\begin{eqnarray}
\label{eq:gammamu}
	&&\Gamma_\mu(q)=g_V(q^2)\gamma_\mu+\ti g_M(q^2)\frac{\sigma_{\mu\nu}}{2m_p}q^\nu\notag\\
	&&~~~~~-g_A(q^2)\gamma_\mu \gamma_5-g_P (q^2)q_\mu \gamma_5,	
\end{eqnarray}
where $\tau_-$ is the isospin lowering operator.
More details about the current operator $\mathcal J_\mu^\dagger$ as well as its nonrelativistic-reduced form can be found in Refs.~\cite{Song2014, Yao2015}.

Here we consider the most probable path for the $0\nu\beta\beta$ decay, namely, the transition between the ground states ($J^\pi=0^+$) of even-even nuclei.
Taking the nuclear wave functions in Eq.~(\ref{eq:GCMwf}) constructed with the GCM+PNAMP (particle-number and angular-momentum projection) method, the total NME reads
\begin{eqnarray}
\label{eq:NME}
	 M^{0\nu} &=& \sum_{\beta_2^I,\beta_2^F} f_{0_F^+}^\ast(\beta_2^F)f_{0_I^+}(\beta_2^I) \int\frac{d^3q}{(2\pi)^3}\, h(q)\notag\\
	 &\times &\sum_{abcd} \langle ab| \,\Gamma_\mu^{(1)}(q)\Gamma^{\mu(2)}(q)\,\te^{\ti\bm q\cdot(\bm x_1-\bm x_2)}\,|cd\rangle \\
	 &\times & \langle \beta_2^F|\,c_a^{(\pi)\dagger}c_b^{(\pi)\dagger}c_d^{(\nu)}c_c^{(\nu)}\,\hat P^{J =0}\hat P^{N_I} \hat P^{Z_I} \, |\beta_2^I\rangle,\notag
\end{eqnarray}
which is a weighted superposition of the projected matrix elements with different initial and final deformation parameters $\beta_2^I$ and $\beta_2^F$.
The neutron annihilation operators $c_{c,d}^{(\nu)}$ and proton creation operators $c_{a,b}^{(\pi)\dagger}$ are responsible for transforming two neutrons into protons.

To take into account the SRCs of two interacting nucleons, the $0\nu\beta\beta$-decay NME are calculated with nuclear wave functions modified by a Jastrow correlation function~\cite{Jastrow1955, Miller1976},
\begin{eqnarray}
\label{eq:Fofr}
	F(r)=1-c\te^{-ar^2}(1-br^2),
\end{eqnarray}
where $r\equiv |\bm x_1-\bm x_2|$ is the distance of two nucleons.
This is equivalent to modifying the decay operator, $\hat{\mathcal O}^{0\nu}(r)\to F(r)\hat{\mathcal O}^{0\nu}(r)F(r)$. 
Therefore, the single integration over $\bm q$ in Eq.~(\ref{eq:NME}) now becomes twofold:
\begin{eqnarray}
\label{eq:NME-SRC}
	&&\int\frac{d^3q}{(2\pi)^3}\, h(q)\,\Gamma_\mu^{(1)}(q)\Gamma^{\mu(2)}(q)\,\te^{\ti\bm q\cdot(\bm x_1-\bm x_2)}  \\
	&& \Rightarrow\int\!\frac{d^3k}{(2\pi)^3}\,\tilde G(k)\int\!\frac{d^3q}{(2\pi)^3}\, h(q)\,\Gamma_\mu^{(1)}(q)\Gamma^{\mu(2)}(q)\,\te^{\ti(\bm q+\bm k)\cdot(\bm x_1-\bm x_2)}.\notag
\end{eqnarray}
Note that the Fourier transform of the correlation function,
\begin{eqnarray}
\label{eq:Gofk}
	\tilde G(k)\equiv\int\! d^3 r\, F^2(r)e^{-i\bm k\cdot\bm r},
\end{eqnarray}
is used to treat the NME in the reciprocal spaces.

\section{Numerical details}\label{numerical}

The single-particle Dirac equation is solved by expanding the wave functions in the three-dimensional harmonic oscillator basis with 12 major shells~\cite{Gambhir1990}.
A zero-range force $V_0^{pp}\delta(\bm r_1-\bm r_2)$ is implemented in the particle-particle channel.
The pairing strength parameters $V_0^{pp}$ are $-314.55~\mathrm{MeV~fm}^3$ for neutrons and $-346.5~\mathrm{MeV~fm}^3$ for protons, determined by reproducing the corresponding pairing gaps of separable finite-range pairing force~\cite{Tian2009} in ${}^{150}$Nd (see Fig.~1 of Ref.~\cite{Song2014}).
Note that only the like-particle pairing has been considered here. The isovector or isoscalar proton-neutron pairing is not included and the isospin symmetry is broken.
On the one hand, the problem with isospin symmetry has been addressed in the QRPA~\cite{Suhonen1993, Simkovic2008} and the IBM calculations~\cite{Barea2013}, respectively.
It is proposed that the (partial) restoration of isospin symmetry can be achieved by imposing the condition that the $2\nu\beta\beta$ Fermi matrix elements $M^{2\nu}_\mathrm{F}$ vanish. 
This has been realized by adjusting the value of the renormalization constant $g_{pp}^{T=1}$ in QRPA~\cite{Simkovic2013, Fang2015} or by modifying the mapped fermion operators in IBM~\cite{Barea2015}.
Although the Fermi matrix elements $M_\mathrm{F}^{0\nu}$ are considerably reduced, the restoration of isospin symmetry has only a limited effect on the total NMEs.
On the other hand, it has been known in the case of QRPA that the effect of the inclusion of the isoscalar pairing is significant.
The renormalization parameter $g_{pp}^{T=0}$ is crucial to the NME calculation, and its value is usually determined by the requirement that the calculated $2\nu\beta\beta$ Gamow-Teller matrix elements $M_\mathrm{GT}^{2\nu}$ reproduce their experimental values~\cite{Rodin2003}.
Recently, this issue has been revisited by taking the isoscalar-pairing amplitude as a generator coordinate in GCM~\cite{Hinohara2014, Menendez2016}.
This effect turns out to quench the NME $M^{0\nu}$ significantly by a factor even larger than $50\%$.
Inclusion of this effect in CDFT is not trivial and is to be investigated as the next step of our study.

The generator coordinates are chosen in the interval of $\beta_2\in[-0.4,0.6]$ with a step size $\Delta\beta_2=0.1$.
The empirical values for the energy denominator $E_d=1.12A^{1/2}~\mathrm{MeV}$ ($E_d\simeq 13.72~\mathrm{MeV}$ for $A=150$), proposed by Haxton \emph{et al.}~\cite{Haxton1984} and examined in Ref.~\cite{Song2014}, are used in the calculations of the NME with light-neutrino exchange.

Three parametrizations for the Jastrow SRC function $F(r)$~\cite{Simkovic2009a, Jastrow1955, Miller1976}---Miller-Spencer (M-S), Argonne V18 (Argonne), and CD~Bonn (Bonn)---are discussed and the final results with the Argonne parameters $a=1.59~\mathrm{fm}^{-2}$, $b=1.45~\mathrm{fm}^{-2}$, and $c=0.94$ are shown.

\section{Results and discussion}\label{result}

\subsection{NME with light- and heavy-neutrino exchange}

We now discuss in detail the NME for the $0\nu\beta\beta$ decay, ${}^{150}$Nd $\to$ ${}^{150}$Sm, mediated by the exchange of light and heavy neutrinos, respectively.

\begin{figure}[!htbp]
	\centerline{
	\includegraphics[width=8.6cm]{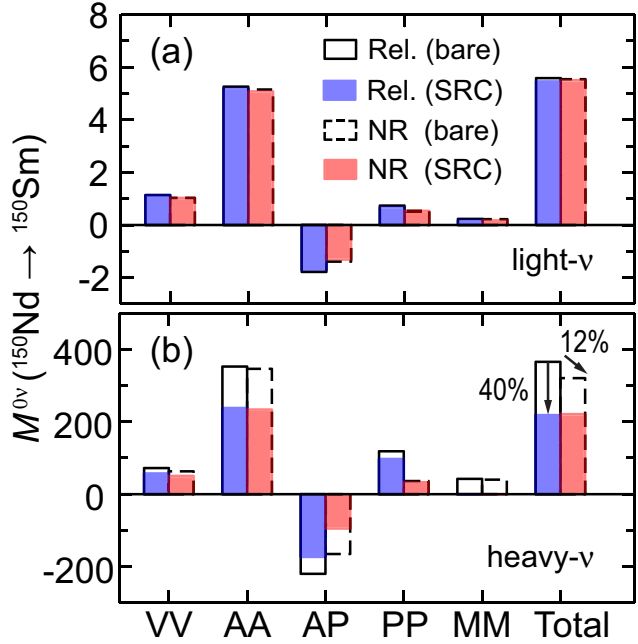}}
	\caption{NME $M^{0\nu}$ for the $0\nu\beta\beta$ decay of ${}^{150}$Nd $\to$ ${}^{150}$Sm mediated by ({a}) light- and ({b}) heavy-neutrino exchange, with the total and the VV, AA, AP, PP, and MM components separately. Results are calculated within the GCM+PNAMP scheme based on the CDFT using both the full relativistic (Rel.) and the nonrelativistic-reduced (NR) decay operators with (SRC) and without (bare) the Argonne-parametrized SRCs.}\label{fig:nme}
\end{figure}

The major results of this paper for the $0\nu\beta\beta$ NME, labeled as ``Rel.~(SRC)'' in Fig.~\ref{fig:nme}, are given by the calculations based on the full relativistic decay operator and the Jastrow SRCs using the Argonne parametrization.
The values for the total NME are $M^{0\nu}=5.46$ in the light-neutrino mechanism and $M^{0\nu}=218.2$ in the heavy-neutrino case.
Furthermore, the results obtained from the relativistic operator and the nonrelativistic-reduced operator are compared side by side (Rel. vs. NR) in the figure.
For each case, two sets of values, obtained with and without considering the SRCs, are distinguished by the color-filled and open bars, respectively.

According to the different coupling channels of $\Gamma_\mu(q)$ in Eq.~(\ref{eq:gammamu}), the total NME can be decomposed into vector (VV), axial-vector (AA), axial-vector and pseudoscalar (AP), pseudoscalar (PP), and weak-magnetism (MM) terms.
Figure~\ref{fig:nme} shows the contributions of these individual terms to the total NMEs in different cases.
All of them are consistent with the conclusion in Ref.~\cite{Yao2015} that the AA term exhausts more than 95\% of the total NME.
The values for the total NMEs are listed in Table~\ref{tab:finalnme}.

\begin{table}[!htbp]
    \caption{NME $M^{0\nu}$ for the $0\nu\beta\beta$ decay of ${}^{150}$Nd $\to$ ${}^{150}$Sm, calculated within the GCM+PNAMP scheme based on the CDFT using both the full relativistic (Rel.) and nonrelativistic-reduced (NR) decay operators with (SRC) and without (bare) the Argonne-parametrized SRCs. The bold data are our recommended values.}\label{tab:finalnme}
    \begin{ruledtabular}
    \begin{tabular}{ccccc}\\[-15pt]
        \multirow{2}{*}{${}^{150}\mathrm{Nd}$} & \multicolumn{2}{c}{NME (light-$\nu$)} & \multicolumn{2}{c}{NME (heavy-$\nu$)} \\[2pt]
        \cline{2-3} \cline{4-5}\\[-14pt]
         &  bare & SRC &  ~bare & SRC \\[1pt]
        \colrule\\[-14pt]
        Rel. & $5.59$ & ${\bf 5.46}$ & ~$365.3$ & ${\bf 218.2}$ \\[1pt]
        NR   & $5.55$ & $5.51$ & ~$320.3$ & $220.8$ \\[1pt]
        \end{tabular}
    \end{ruledtabular}
\end{table}

Comparing to our previous calculations for the light-neutrino NME~\cite{Song2014}, the new results obtained here after implementing the SRCs indicate that the SRC effects can be safely neglected in this circumstance.
Moreover, the calculation confirms our previous conclusion that the nonrelativistic reduction of the decay operator is a very good approximation to the full operator in the light-neutrino NME, regardless of whether the SRCs are included.

The heavy-neutrino NME, however, has a more sensitive response to both the inclusion of the SRCs and the nonrelativistic reduction of the decay operator.
First, Fig.~\ref{fig:nme}({b}) shows that the SRCs introduce a significant reduction in the total NME up to 40\%.
This can be understood by considering the short-range nature of the heavy-neutrino exchange process, as we shall see in the detailed investigation later.
Second, the impacts of relativity on the heavy-neutrino NME manifest clearly a dual feature; while the nonrelativistic approximation results in a reduction of 12\% in the bare NME, this effect is completely compensated after the implementation of the SRCs.
The cancellation of relativistic effects mainly comes from the PP and AP channels whose contributions have the opposite signs.
With the onset of the SRCs, the positive relativistic effects in the PP channel are decreased while the magnitude of the negative relativistic effects in the AP channel are increased, resulting in the final elimination of the difference in the total NME.
The interplay between the effects of SRCs and relativity in the heavy-neutrino NME will be further discussed in the following.

\subsection{Effects of SRCs}

The disparate SRC responses of the light- and heavy-neutrino $0\nu\beta\beta$ NME can be well understood by decomposing the NME into its contributions from the various channels $i=$VV, AA, AP, PP, and MM. For this purpose, we rewrite the NME in Eq.~(\ref{eq:NME}) as
\begin{eqnarray}
\label{eq:NMEq}
	 M^{0\nu}_i \equiv \frac{4\pi R}{g_A^2}\int \frac{q^2 dq}{(2\pi)^3}\, H_i(q)I_i(q).
\end{eqnarray}
Here the $q$ dependence in $\Gamma_\mu(q)$ is put into the function $H_i(q)$, i.e.,
\begin{subequations}
\label{Hofqs}
	\begin{eqnarray}
		H_\mathrm{VV}(q)&=& h(q)\,g_V^2(q^2)\,, \\
		H_\mathrm{AA}(q)&=& h(q)\,g_A^2(q^2)\,, \\
		H_\mathrm{AP}(q)&=& h(q)\,g_A(q^2)g_P(q^2)q\,, \\
		H_\mathrm{PP}(q)&=& h(q)\,g_P^2(q^2)q^2\,, \\
		H_\mathrm{MM}(q)&=& h(q)\,g_M^2(q^2)q^2/4m_p^2\,.
	\end{eqnarray}
\end{subequations}
For simplicity, the other parts of the NME in Eq.~(\ref{eq:NME}) that are not included in $H_i(q)$ are defined as a new function $I_i(q)$, which is also channel specified and $q$ dependent.
With this definition, the SRC-corrected NME, which contains a twofold integration as in Eq.~(\ref{eq:NME-SRC}), can be calculated by simply replacing $H_i(q)$ with a modified function $H_{i}^\mathrm{src}(q)$ in Eq.~(\ref{eq:NMEq}),
\begin{eqnarray}
\label{eq:Hsrc}
	H_{i}^\mathrm{src}(q)=H_i(q)+\int\frac{q^{\prime 2}dq^\prime}{(2\pi)^2}H_i(q^\prime)\frac{1}{2qq^\prime}\int_{(q-q^\prime)^2}^{(q+q^\prime)^2}du \,g(u),\notag\\
\end{eqnarray}
where $g(u)= 4\pi\int_0^\infty \left[F^2(r)-1\right]j_0(kr)r^2dr$, where $u\equiv k^2$, $F(r)$ is the aforementioned Jastrow SRC correlation function, and $j_0(kr)$ is the spherical Bessel function.

\begin{figure}[!htbp]
	\centerline{
	\includegraphics[width=8.6cm]{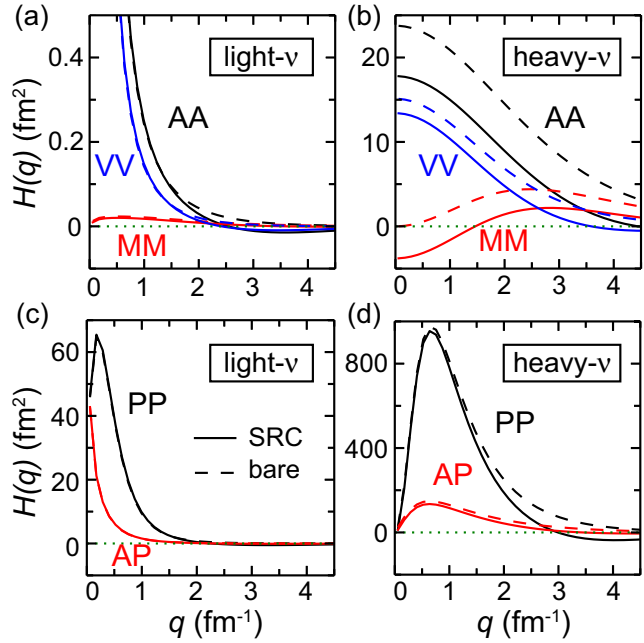}}
	\caption{The function $H_i(q)$ with (SRC) and without (bare) the Argonne-SRC modification for the VV, AA, AP, PP and MM channels in the $0\nu\beta\beta$ NME of light- and heavy-neutrino exchange, respectively.}\label{fig:src}
\end{figure}

The information regarding the decay mechanism of light- or heavy-neutrino exchange is contained exclusively in the function $H_i(q)$ in  Eq.~(\ref{eq:NMEq}) or, in $H_{i}^\mathrm{src}(q)$ after the modification with the SRCs.
Figure~\ref{fig:src} shows the function $H_i(q)$ (bare) in comparison with the SRC-modified function $H_{i}^\mathrm{src}(q)$ (SRC) for the light- and heavy-neutrino cases, respectively.
For heavy-neutrino exchange [Figs.~\ref{fig:src}({b}) and ~\ref{fig:src}({d})], the $H(q)$ functions are altered significantly by the SRC correction.
For instance, the downward shift of $H_\mathrm{AA}(q)$ is responsible for the large-amplitude reduction of the AA matrix element by the SRCs in Fig.~\ref{fig:nme}({b}).
The curve of $H_\mathrm{MM}(q)$ is also shifted downward.
In this case, it becomes negative in the low-$q$ range, leading to a cancellation of the SRC-corrected MM matrix element after the $q$ integration.
On the contrary, Figs.~\ref{fig:src}({a}) and ~\ref{fig:src}({c}) show only minor differences between the $H(q)$ functions with and without including the SRC correction in different channels of the light-neutrino NME.
This explains the reason why the light-neutrino NMEs are merely affected by the SRCs, and can be easily interpreted in terms of the $q$ dependence of neutrino potential $h(q)$.
Unlike the constant $h(q)$ in Eq.~(\ref{eq:hofq-heavy}) for heavy-neutrinos, the light-neutrino $h(q)$ in Eq.~(\ref{eq:hofq-light}) grows very sharply when $q\to 0$ and vanishes very rapidly as $q$ increases.
As a result, for light neutrinos, the $h(q)$ dominates the $q$ dependence of the $H(q)$ function, diminishing the difference between $H_i(q)$ and $H^\mathrm{src}_i(q)$.
Therefore, the differences in the $q$ dependence of the neutrino potential $h(q)$ cause the different effects that the SRCs have on the light- and heavy-neutrino NMEs.

To validate the above conclusions in a systematic way, we generalize the NME calculations to several other $0\nu\beta\beta$ candidate nuclei, by considering three parametrizations for the SRC function $F(r)$ in Eq.~(\ref{eq:Fofr}): M-S, Argonne, and Bonn, using the parameters determined in Refs.~\cite{Miller1976, Simkovic2009a, Jastrow1955}.
The systematic calculations are performed with the full relativistic decay operator and the particle-number projected spherical wave functions, where the normalized NMEs are provided as
\begin{eqnarray}
\label{eq:NME-sph}
	M^{0\nu}_\mathrm{sph}&=&\dfrac{\langle \beta_2^F=0\vert \hat{\mathcal O}^{0\nu}\hat P^{N_I}\hat P^{Z_I}\vert \beta_2^I=0\rangle}
{ \prod_{a=I, F}\sqrt{\langle \beta_2^a=0| \hat P^{N_a}\hat P^{Z_a}  | \beta_2^a=0\rangle}}\;.
\end{eqnarray}

\begin{table*}[!htbp]
    \caption{Normalized NME $M^{0\nu}_\mathrm{sph}$ for the $0\nu\beta\beta$ decay obtained
with the particle-number projected spherical mean-field configurations ($\beta_2^I=\beta_2^F=0$) based on CDFT. Columns 2--8 list the calculated light-neutrino NME without (bare) and with three types of SRC, respectively. Columns 9--15 show the counterparts in the case of heavy neutrinos. Also shown are the relative corrections $\Delta_\mathrm{src}$.}\label{tab:src}
    \begin{ruledtabular}
    \begin{tabular}{rcccccccccccccc}\\[-15pt]
         & \multicolumn{7}{c}{NME (light-$\nu$)} & \multicolumn{7}{c}{NME (heavy-$\nu$)} \\[2pt]
        \cline{2-8} \cline{9-15}\\[-14pt]
         &  bare & M-S & $\Delta_\mathrm{src}$ & {Argonne} & $\Delta_\mathrm{src}$ & {Bonn} & $\Delta_\mathrm{src}$
         & ~bare & M-S & $\Delta_\mathrm{src}$ & {Argonne} & $\Delta_\mathrm{src}$ & {Bonn} & $\Delta_\mathrm{src}$ \\[1pt]
        \colrule\\[-14pt]
         ${}^{48}$Ca~ &  $3.67$  & $3.26$   & $11\%$ &  $3.62$   & $ 1\%$ &  $3.74$   &$-2\%$~
                      & ~$145.6$ &~$42.8$ ~ & $71\%$ & ~$82.3$ ~ & $43\%$ & ~$117.0$~ &$20\%$ \\[1pt]
         ${}^{76}$Ge~ &  $7.61$  & $6.36$   & $17\%$ &  $7.48$   & $ 2\%$ &  $7.84$   &$-3\%$~
                      & ~$466.8$ &~$135.7$~ & $71\%$ & ~$267.0$~ & $43\%$ & ~$378.1$~ &$19\%$\\[1pt]
         ${}^{82}$Se~ &  $7.60$  & $6.38$   & $16\%$ &  $7.48$   & $ 2\%$ &  $7.83$   &$-3\%$~
                      & ~$454.0$ &~$132.7$~ & $71\%$ & ~$261.4$~ & $42\%$ & ~$369.0$~ &$19\%$ \\[1pt]
         ${}^{96}$Zr~ &  $5.68$  & $4.84$   & $15\%$ &  $5.58$   & $ 2\%$ &  $5.82$   &$-2\%$~
                      & ~$307.3$ &~$89.0$ ~ & $71\%$ & ~$177.7$~ & $42\%$ & ~$250.5$~ &$18\%$\\[1pt]
        ${}^{100}$Mo~ &  $10.99$ & $9.38$   & $15\%$ &  $10.80$  & $ 2\%$ &  $11.27$  &$-3\%$~
                      & ~$596.3$ &~$174.1$~ & $71\%$ & ~$346.7$~ & $42\%$ & ~$487.4$~ &$18\%$\\[1pt]
        ${}^{116}$Cd~ &  $6.19$  & $5.18$   & $16\%$ &  $6.08$   & $ 2\%$ &  $6.37$   &$-3\%$~
                      & ~$378.3$ &~$111.3$~ & $71\%$ & ~$222.7$~ & $41\%$ & ~$311.2$~ &$18\%$\\[1pt]
        ${}^{124}$Sn~ &  $6.70$  & $5.68$   & $15\%$ &  $6.58$   & $ 2\%$ &  $6.87$   &$-3\%$~
                      & ~$381.2$ &~$111.7$~ & $71\%$ & ~$224.6$~ & $41\%$ & ~$313.8$~ &$18\%$\\[1pt]
        ${}^{130}$Te~ &  $9.55$  & $8.03$   & $16\%$ &  $9.38$   & $ 2\%$ &  $9.82$  &$-3\%$~
                      & ~$573.0$ &~$168.5$~ & $71\%$ & ~$339.2$~ & $41\%$ & ~$472.8$~ &$17\%$\\[1pt]
        ${}^{136}$Xe~ &  $6.62$  & $5.58$   & $16\%$ &  $6.51$   & $ 2\%$ &  $6.80$   &$-3\%$~
                      & ~$394.5$ &~$116.3$~ & $71\%$ & ~$234.3$~ & $41\%$ & ~$326.2$~ &$17\%$\\[1pt]
        ${}^{150}$Nd~ &  $13.26$ & $11.11$  & $16\%$ &  $13.00$  & $ 2\%$ &  $13.62$  &$-3\%$~
                      & ~$804.1$ &~$237.7$~ & $70\%$ & ~$481.7$~ & $40\%$ & ~$667.9$~ &$17\%$\\[1pt]
        \end{tabular}
    \end{ruledtabular}
\end{table*}

Table~\ref{tab:src} shows the calculated NMEs $M^{0\nu}_\mathrm{sph}$ for ten candidate nuclei, ranging from ${}^{48}$Ca to ${}^{150}$Nd, for the $0\nu\beta\beta$ decay mediated by light- and heavy-neutrino exchange.
The relative corrections $\Delta_\mathrm{src}\equiv (M^{0\nu}_\mathrm{bare}-M^{0\nu}_\mathrm{src})/M^{0\nu}_\mathrm{bare}$ represent the SRC effects in a quantitative way.
Columns 2--8 of Table~\ref{tab:src} list the calculated light-neutrino NMEs without (bare) and with three types of SRC, as well as the relative corrections $\Delta_\mathrm{src}$.
Columns 9--15 show the counterparts in the case of heavy neutrinos.

Consistent with the full GCM calculation for ${}^{150}$Nd, the inclusion of the Argonne-parametrized SRCs can reduce the light-neutrino NME by a factor of 1\%--3\% and the heavy-neutrino NME by a factor of 40\%--44\%.
In the case of light-neutrinos, only the M-S SRCs provide a noticeable correction of about $15\%$.
Both the Argonne and the Bonn SRCs have few influences on the total NME.
For the heavy-neutrino NME, the M-S and the Bonn SRCs introduce the most significant (70\%) and the most modest (15\%--20\%) quenching effects, respectively.
The correction given by the Argonne parametrization lies in between.

In the calculation of the $0\nu\beta\beta$ NMEs for the heavy-neutrino exchange mode, it is not surprising that the short-range effects play a significant role.
Besides the nucleon-nucleon SRCs, the effect of finite nucleon size (FNS) also comes into play.
The FNS effect is considered in this work by employing the phenomenological dipole nucleon form factors in the momentum space~\cite{Ericson1988, Kitagaki1983}.
The sensitivity of the heavy-neutrino NMEs to the form factors has been manifested in Ref.~\cite{Simkovic1992} via the calculation with both the phenomenological form factors and the form factors deduced from the quark confinement model.
Despite that there exist only small differences between the two types of the form factors, the resulting values for the NMEs differ by almost one order of magnitude.
Furthermore, it is seen for the heavy-neutrino exchange mode in Ref.~\cite{Simkovic1992} as well as in this paper that the absolute values of $M^{0\nu}_\mathrm{AP}$ and $M^{0\nu}_\mathrm{PP}$, which are originated from the nucleon pseudoscalar coupling interaction, are comparable in size to that of $M^{0\nu}_\mathrm{AA}$ and $M^{0\nu}_\mathrm{VV}$.
This fact, according to Ref.~\cite{Simkovic1992}, emphasizes the importance of the alternative $0\nu\beta\beta$-decay mechanisms such as double charge exchange of the pions in flight between the two nucleons~\cite{Vergados1982}.
A similar conclusion has also been drawn in the framework of $R$-parity-violating supersymmetry that the pion-exchange mechanism may dominate over the conventional two-nucleon one if the $0\nu\beta\beta$ decay is mediated by heavy neutrinos~\cite{Faessler1997, Prezeau2003}. 
Thus, it still needs more investigations as to the accurate treatment of the FNS as well as the $0\nu\beta\beta$-decay mechanisms in the calculation with heavy neutrinos.

\subsection{Effects of relativity}

The relativistic correction that is missing in the nonrelativistic approximation is of the order of $(q/m_p)^4$ at the lowest level.
In other words, the effects of this correction display a high-$q$ character.
Consequently, relativity does not play an important role in the calculation of light-neutrino NME owing to the large suppression of $h(q)$ in the intermediate- and high-$q$ regions.
There are small differences in the individual channels, especially the PP and the AP channels, but the differences almost cancel out in the total NME.

\begin{figure*}[!htbp]
	\centerline{
	\includegraphics[width=12cm]{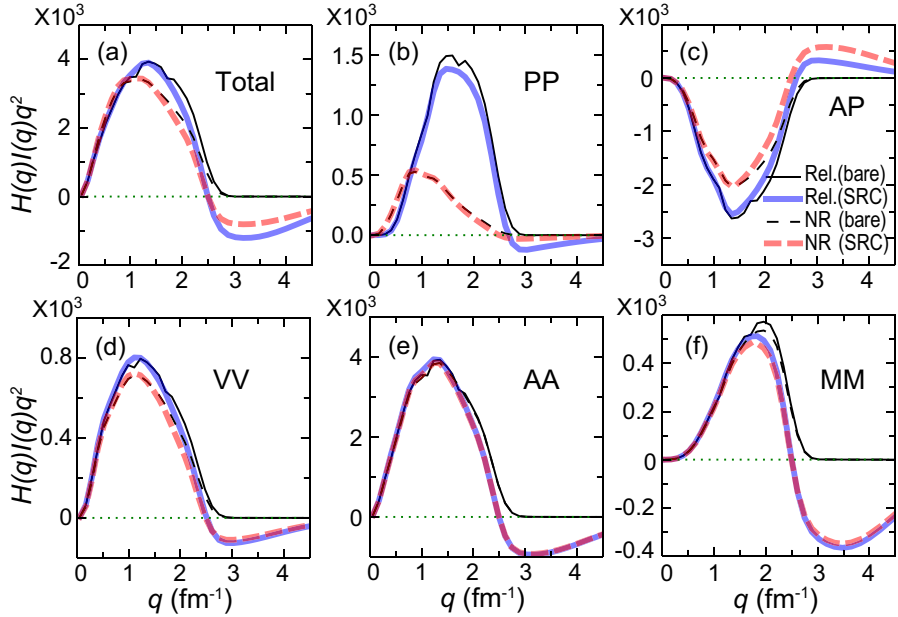}}
	\caption{The $q$-space distribution of the $0\nu\beta\beta$ NME with heavy-neutrino exchange. Comparisons are made between the calculations using both the full relativistic (Rel.) and the nonrelativistic-reduced (NR) decay operators with (SRC) and without (bare) the Argonne-parametrized SRCs. Particle-number projected spherical wave functions are used  in this calculation for the initial nucleus ${}^{150}$Nd and the final nucleus ${}^{150}$Sm.}\label{fig:rel}
\end{figure*}

For the heavy-neutrino NME, the relativistic corrections have a more significant effect.
As we have seen in Fig.~\ref{fig:nme}({b}), the contribution of the relativistic correction constitutes about 12\% of the total NME without switching on the SRCs.
In this case the effects in the PP and other positive terms are not entirely canceled out by the negative contribution arising from the AP term.
With the SRCs, however, the positive and negative contributions from those individual terms become compensated with each other as in the light-neutrino case.
So there are no remarkable effects left in the total NME.

The cancellation mainly comes from the PP and AP channels.
Figure~\ref{fig:rel}({b}) shows for the PP channel that the $q$-space distribution of the heavy-neutrino NME, $H_i(q)I_i(q)q^2, i=\mathrm{PP}$, is only modified slightly by the SRCs when the nonrelativistic operator is used [``NR~(bare)'' vs ``NR~(SRC)''], while the distribution changes remarkably in the full relativistic calculation [``Rel.~(bare)'' vs ``Rel.~(SRC)''].
As a result, the nonrelativisitc NME in the PP channel is almost unchanged by the SRCs, while the relativistic NME gets reduced, leading to a smaller difference between the two NMEs, i.e., a relatively weak relativistic effect.
The opposite is found for the AP channel in Fig.~\ref{fig:rel}({c}).
The relativistic effects in this channel are enhanced by the SRCs as the nonrelativisitc curve is modified more significantly.
The other channels, whose $q$-space distributions are shown in Figs.~\ref{fig:rel}({d})--\ref{fig:rel}({f}), have little contribution to the relativistic effects.
Notably, the relativistic corrections in the AP and PP channels have the opposite signs.
Therefore, the decrease of the positive contribution and the increase of the negative term diminish the overall (positive) effects that appear in the bare NME.
From the $q$-space distribution of the total NME, $\sum_i H_i(q)I_i(q)q^2$, shown in Fig.~\ref{fig:rel}({a}), it is also clearly seen that the SRCs affect the relativistic NME more significantly than the nonrelativistic one, resulting in an overall reduction of the relativistic effect.
For the sake of simplicity, the functions plotted in Fig.~\ref{fig:rel} are extracted from the NME-calculations with only spherical configurations of the initial and final nuclear states.
The features we discuss here should apply to the complete GCM calculations without loss of generality.

We have carried out systematic investigations of the relativistic effects on the $0\nu\beta\beta$-decay NMEs of other candidate nuclei.
The normalized NMEs of Eq.~(\ref{eq:NME-sph}) are calculated with the relativistic and nonrelativistic operators respectively, and the relative corrections $\Delta_\mathrm{Rel.}\equiv (M^{0\nu}_\mathrm{Rel.}-M^{0\nu}_\mathrm{NR})/M^{0\nu}_\mathrm{Rel.}$ are extracted.

\begin{table}[!htbp]
    \caption{Relativistic correction $\Delta_\mathrm{Rel.}$ in the $0\nu\beta\beta$-decay NME with (SRC) and without (bare) the Argonne-parametrized SRCs. Particle-number projected spherical mean-field wave functions ($\beta_2^I=\beta_2^F=0$) based on the CDFT are used in the calculation.}\label{tab:rel}
    \begin{ruledtabular}
    \begin{tabular}{rrcrc}\\[-15pt]
         & \multicolumn{2}{c}{$\Delta_\mathrm{Rel.}$ (light-$\nu$)} & \multicolumn{2}{c}{$\Delta_\mathrm{Rel.}$ (heavy-$\nu$)} \\[2pt]
        \cline{2-3} \cline{4-5}\\[-14pt]
         &  bare & SRC &  ~bare & SRC \\[1pt]
        \colrule\\[-14pt]
         ${}^{48}$Ca~ & $-2\%$ & $-1\%$ & ~$15\%$ & $-2\%$ \\[1pt]
         ${}^{76}$Ge~ & $-1\%$ & $-3\%$ & ~$10\%$ & $-6\%$ \\[1pt]
         ${}^{82}$Se~ & $-1\%$ & $-3\%$ & ~$11\%$ & $-5\%$ \\[1pt]
         ${}^{96}$Zr~ & $ 1\%$ & $-1\%$ & ~$11\%$ & $-2\%$ \\[1pt]
        ${}^{100}$Mo~ & $ 1\%$ & $-1\%$ & ~$11\%$ & $-2\%$ \\[1pt]
        ${}^{116}$Cd~ & $ 1\%$ & $-1\%$ & ~$12\%$ & $-3\%$ \\[1pt]
        ${}^{124}$Sn~ & $-1\%$ & $-2\%$ & ~$10\%$ & $-3\%$ \\[1pt]
        ${}^{130}$Te~ & $-1\%$ & $-2\%$ & ~$10\%$ & $-3\%$ \\[1pt]
        ${}^{136}$Xe~ & $-1\%$ & $-3\%$ & ~$10\%$ & $-3\%$ \\[1pt]
        ${}^{150}$Nd~ & $ 1\%$ & $-0\%$ & ~$13\%$ & $-0\%$ \\[1pt]
        \end{tabular}
    \end{ruledtabular}
\end{table}

Shown in Table~\ref{tab:rel} are the values of $\Delta_\mathrm{Rel.}$ obtained for both the light- and heavy-neutrino exchange NMEs with and without considering the SRC effects.
Consistent with the full GCM calculation for ${}^{150}$Nd, the error arisen from the nonrelativistic approximation for the light-neutrino NME is marginal.
It increases or decreases the total NME by a factor less than 5\%.
The relativistic corrections become more significant in the heavy-neutrino case where we find that the nonrelativistic calculations underestimate the bare NME by 10\%--15\% while they overestimate the SRC-corrected NME by a factor of roughly 5\%.
Interestingly, the SRCs, by affecting the PP and AP channels differently, not only reduce the relativistic effects observed in the bare NMEs, but also reverse the signs of net effects in most circumstances.

\subsection{Comparison and discussion}
\label{discussion}

Table~\ref{tab:compare} displays our final NMEs for the $0\nu\beta\beta$ decay of ${}^{150}\mathrm{Nd}\to {}^{150}\mathrm{Sm}$ in comparison with those from earlier investigations: nonrelativistic EDF~\cite{Rodriguez2010,Vaquero2013}, PHFB~\cite{Rath2012, Rath2013}, QRPA by the T\"ubingen group (QRPA-T\"u)~\cite{Fang2015}, Skyrme QRPA by the North-Carolina group (QRPA-NC)~\cite{Mustonen2013}, and IBM~\cite{Barea2015}.
Here, only the results obtained with consideration of nuclear deformations are adopted for comparison.
All results are calculated with an unquenched axial-vector coupling constant $g_A= 1.254$ or a value close to it and using the radius parameter $R=1.2A^{1/3}~\mathrm{fm}$.

The Argonne parametrization is applied in our calculation for the nucleon-nucleon SRCs, as well as in the listed results of PHFB and IBM.
The nonrelativistic EDF calculation considers the SRCs via the unitary correlation operator method, which, according to Ref.~\cite{Barea2013}, gives similar effects as the Argonne-parametrized Jastrow function.
The QRPA-T\"u calculation uses the Bonn parametrization for the SRCs, while the QRPA-NC calculation neglects the SRCs completely, both of which are expected to result in a larger total NME than the Argonne parametrization.
However, according to Table~\ref{tab:src}, the discrepancies are negligible in the light-neutrino NME.
Hence, the possible uncertainties arisen from different ways of treating the SRCs will not alter the conclusions of this comparison.

\begin{table*}[!htbp]
    \caption{NME for the $0\nu\beta\beta$ decay of ${}^{150}$Nd $\to$ ${}^{150}$Sm mediated by light- and heavy-neutrino exchange. The column ``CDFT'' shows the results of this work in bold face, which are calculated within the GCM+PNAMP scheme based on the CDFT, in comparison with the results from other model calculations. With the latest data for the half-life $T^{0\nu}_{1/2}>2.0\times 10^{22}$ yr (90\% C.L.)~\cite{Arnold2016} and the calculated phase-space factor $G_{0\nu}=63.03\times 10^{-15}~\mathrm{yr}^{-1}$~\cite{Kotila2012}, the limits for the effective neutrino masses $|\langle m_\nu\rangle|$(eV) and $|\langle m_{\nu_h}^{-1}\rangle|^{-1}$($\times 10^6$ GeV) are derived for each model calculation using Eq.~(\ref{eq:half-life}).}\label{tab:compare}
    \begin{ruledtabular}
    \begin{tabular}{ccccccr}\\[-14pt]
        & CDFT & EDF & PHFB & QRPA-T\"u & QRPA-NC & IBM \\[2pt]
        \colrule\\[-14pt]
        light-$\nu$ NME & $\mathbf{5.46}$  & $1.71$/$2.19$ & $2.49$--$3.31$ & $3.37$ & $3.14$/$2.71$ & $2.67$ \\[1pt]
         $|\langle m_\nu\rangle|$  & $\mathbf{< 1.7}$  & $< 5.4$/$4.2$ & $< 3.7$--$2.8$ & $< 2.7$ & $< 2.9$/$3.4$ & $< 3.4$ \\[5pt]
        heavy-$\nu$ NME & $\mathbf{218.2}$ & -- & $77.3$--$97.8$ & -- & --  & $116.0$ \\[1pt]
         $|\langle m_{\nu_h}^{-1}\rangle|^{-1}$ & $\mathbf{> 11.4}$ & -- & $> 4.0$--$5.1$ & -- & --  & $> 6.1$ \\[1pt]
    \end{tabular}
    \end{ruledtabular}
\end{table*}

The EDF calculations are carried out within a similar beyond-mean-field framework as ours and based on the nonrelativistic Gogny functional D1S.
By choosing the quadrupole deformation $\beta_2$ as the generator coordinate in the GCM method, the final NME includes the shape mixing effect and the resulting NME is $M^{0\nu}=1.71$~\cite{Rodriguez2010}.
This value increases to $M^{0\nu}=2.19$ when the pairing fluctuations are included explicitly~\cite{Vaquero2013}.
The results from the PHFB model are obtained with a pairing plus quadrupole-quadrupole interaction, and the ranges presented in the table are given by choosing a series of different parametrizations for this interaction~\cite{Rath2012, Rath2013}.
The NME of QRPA-T\"u is obtained by deformed QRPA calculations based on a set of Woods-Saxon single-particle levels and using the $G$ matrix of the realistic CD~Bonn potential as residual interaction.
Isospin symmetry is partially restored by enforcing the Fermi matrix element $M_\mathrm{F}^{2\nu}=0$~\cite{Fang2015}.
In the QRPA-NC calculations, modern Skyrme functionals (SkM*/modified SkM*) are used in a self-consistent way for generating both the HFB mean fields and the residual interactions in QRPA~\cite{Mustonen2013}.
The IBM results are calculated by applying the interacting boson model IBM-2~\cite{Barea2015}.

Among different nuclear models, our CDFT beyond-mean-field calculation provides the largest values for the NMEs of the $0\nu\beta\beta$ decay for ${}^{150}\mathrm{Nd}\to{}^{150}\mathrm{Sm}$.
In particular, our result obtained for the light-neutrino NME is almost 3 times as large as that of the density-functional method using the nonrelativistic Gogny functional D1S for possible reasons that have been discussed in detail in Refs.~\cite{Song2014, Yao2015}.
Other nuclear models provide predictions for the NME that lie between the two density-functional results.
For the heavy-neutrino mediated $0\nu\beta\beta$ process, the NME is not provided by nonrelativisitc EDF, but our result is larger by a factor of 2 than those from PHFB and IBM.
Moreover, we find that the ratios of the heavy-neutrino NME to the light-neutrino NME given by our calculations and by IBM are surprisingly similar, which are around 40, while the PHFB calculations lead to a smaller ratio of around 30.

The results of double-$\beta$-decay experiments, recently released by the NEMO-3 Collaboration, have set a lower limit of $T_{1/2}^{0\nu}>2.0\times 10^{22}$ yr (90\% C.L.) for the half-life of ${}^{150}$Nd~\cite{Arnold2016}.
With the computed phase-space factor $G_{0\nu}=63.03\times 10^{-15}~\mathrm{yr}^{-1}$~\cite{Kotila2012}, it is straightforward to derive the constraints on the fundamental parameters in $f(m_i,U_{ei})$ according to Eq.~(\ref{eq:half-life}).
Combining the experimental data and the CDFT results for the NMEs, our predictions for the limits of neutrino masses are $|\langle m_\nu\rangle|<1.7~\mathrm{eV}$ for light neutrinos and $|\langle m_{\nu_h}^{-1}\rangle|^{-1}>11.4\times 10^6~\mathrm{GeV}$ for heavy neutrinos.
The predictions by other nuclear models are shown in Table~\ref{tab:compare}.
By comparison, the CDFT beyond-mean-field results impose the most stringent constraints on the effective masses of both light and heavy neutrinos.

\begin{table}[!htbp]
    \caption{The NMEs $M^{0\nu}$ and the limits imposed the effective neutrino masses $|\langle m_\nu\rangle|$ (eV) and $|\langle m_{\nu_h}^{-1}\rangle|^{-1}$ ($\times 10^6$ GeV) based on the present CDFT calculation. The lower limits of the half-life $T^{0\nu}_{1/2} (\times 10^{22}~\text{yr, 90\% C.L.})$ for the $0\nu\beta\beta$ decay are from the most recent measurements~\cite{Umehara2008, Agostini2013, Barabash2011b, Argyriades2010, Arnold2014, Danevich2003, Hwang2009, Barabash2011, Andreotti2011, Gando2016, Arnold2016}, and the phase-space factors $G_{0\nu}(\times 10^{-15}~\mathrm{yr}^{-1})$ are from Ref.~\cite{Kotila2012}.}\label{tab:systematics}
    \begin{ruledtabular}
    \begin{tabular}{rrrllll}\\[-15pt]
    	& $T^{0\nu}_{1/2}$ & ~$G_{0\nu}$~~ & \multicolumn{2}{c}{light-$\nu$} & \multicolumn{2}{c}{heavy-$\nu$} \\[2pt]
        \cline{4-5} \cline{6-7}\\[-14pt]
         && &  ~$M^{0\nu}$ & ~~$|\langle m_\nu\rangle|$ &  ~~$M^{0\nu}$ & ~$|\langle m_{\nu_h}^{-1}\rangle|^{-1}$ \\[2pt]
        \colrule\\[-14pt]
         ${}^{48}$Ca~ & $5.8$  & ~$24.81$~ & ~$2.71$ & ~~$<3.2$ & ~~$84.5$ & ~~~$>4.7$ \\[1pt]
         ${}^{76}$Ge~ & $3000$ & ~$2.363$~ & ~$6.04$ & ~~$<0.2$ & ~~$209.1$ & ~~~$>82.1$ \\[1pt]
         ${}^{82}$Se~ & $36$   & ~$10.16$~ & ~$5.30$ & ~~$<1.0$ & ~~$189.3$ & ~~~$>16.9$ \\[1pt]
         ${}^{96}$Zr~ & $0.92$ & ~$20.58$~ & ~$6.37$ & ~~$<3.7$ & ~~$220.9$ & ~~~$>4.5$ \\[1pt]
        ${}^{100}$Mo~ & $110$  & ~$15.92$~ & ~$6.48$ & ~~$<0.4$ & ~~$232.6$ & ~~~$>45.4$ \\[1pt]
        ${}^{116}$Cd~ & $17$   & ~$16.70$~ & ~$5.43$ & ~~$<1.1$ & ~~$201.1$ & ~~~$>15.8$ \\[1pt]
       ${}^{124}$Sn~ & $0.005$ & ~$9.04$~  & ~$4.25$ & ~~$<114$ & ~~$168.5$ & ~~~$>0.2$ \\[1pt]
        ${}^{130}$Te~ & $280$  & ~$14.22$~ & ~$4.89$ & ~~$<0.3$ & ~~$193.8$ & ~~~$>57.1$ \\[1pt]
       ${}^{136}$Xe~ & $10700$ & ~$14.58$~ & ~$4.24$ & ~~$<0.06$ & ~~$166.3$ & ~~~$>306.5$ \\[1pt]
        ${}^{150}$Nd~ & $2.0$  & ~$63.03$~ & ~$5.46$ & ~~$<1.7$ & ~~$218.2$ & ~~~$>11.4$ \\
        \end{tabular}
    \end{ruledtabular}
\end{table}

\begin{figure}[!htbp]
	\centerline{
	\includegraphics[width=8.6cm]{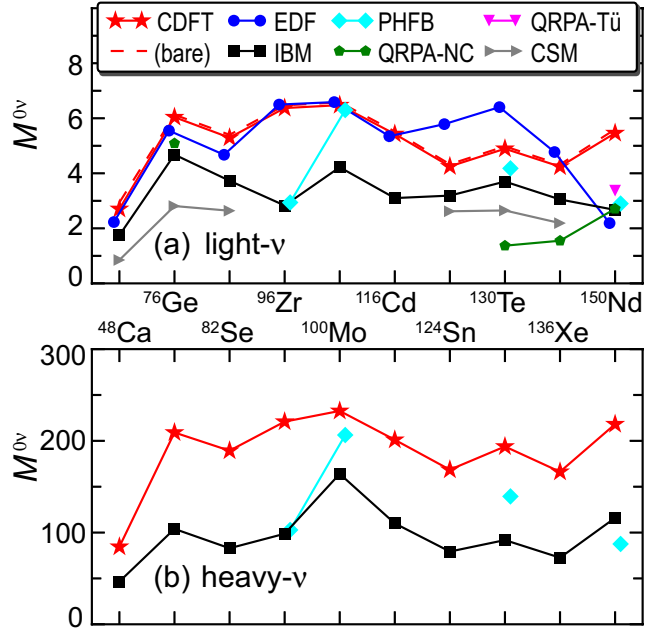}}
	\caption{Comparison of the NMEs $M^{0\nu}$ of the $0\nu\beta\beta$ decay from different model calculations, which include the EDF~\cite{Vaquero2013}, IBM~\cite{Barea2015}, PHFB~\cite{Rath2012, Rath2013}, QRPA-NC~\cite{Mustonen2013}, QRPA-T\"u~\cite{Fang2015}, and CSM~\cite{Menendez2009} calculations, as well as the CDFT calculation in this paper with the GCM+PNAMP wave functions and the Argonne-parametrized SRCs. The CDFT results without considering the SRC effect~\cite{Yao2015} is also shown for the light-neutrino exchange mode by the dashed line in panel (a).}\label{fig:allnme}
\end{figure}

TABLE~\ref{tab:systematics} lists our final NMEs $M^{0\nu}$ of the $0\nu\beta\beta$ decay in ten candidate nuclei for both the light- and the heavy-neutrino exchange modes.
According to the lower limits of the half-life $T^{0\nu}_{1/2}$ from the most recent measurements~\cite{Umehara2008, Agostini2013, Barabash2011b, Argyriades2010, Arnold2014, Danevich2003, Hwang2009, Barabash2011, Andreotti2011, Gando2016, Arnold2016} and the phase-space factors $G_{0\nu}$~\cite{Kotila2012}, the limits on the effective neutrino masses $|\langle m_\nu\rangle|$ and $|\langle m_{\nu_h}^{-1}\rangle|^{-1}$ are further estimated, respectively.
So far, the most stringent constraints are set by the case of ${}^{136}$Xe, which implies that $|\langle m_\nu\rangle|<0.06~\mathrm{eV}$ for light neutrinos and $|\langle m_{\nu_h}^{-1}\rangle|^{-1}>3.065\times 10^8~\mathrm{GeV}$ for heavy neutrinos.
Finally, the CDFT results are compared with the NMEs $M^{0\nu}$ recently obtained from other nuclear models in Fig.~\ref{fig:allnme}.
Our results are among the largest values of the existing calculations in most cases, except that the NMEs $M^{0\nu}$ for ${}^{124}$Sn and ${}^{130}$Te are considerably smaller than those given by the nonrelativistic EDF calculation. 
The agreements with the EDF results are remarkable in the nuclei other than ${}^{124}$Sn, ${}^{130}$Te, and ${}^{150}$Nd.

\section{Summary}\label{summary}

The $0\nu\beta\beta$-decay NMEs have been calculated within the framework of beyond-mean-field CDFT by considering the underlying mechanisms of both light- and heavy-neutrino exchange.
In particular, by investigating in detail the effects of relativity and SRCs in ${}^{150}$Nd, we come to the following conclusions.
(1) Both effects are negligible for the light-neutrino NME, which indicates that the nonrelativistic reduction to the decay operator is a good approximation and the SRC correction can be safely neglected.
(2) The heavy-neutrino NME is more sensitive to both the relativistic correction and the inclusion of SRC than in the light-neutrino case. Therefore, it should be treated more carefully.
(3) For the SRCs, the M-S and the Bonn parametrizations, respectively, introduce the most and the least quenching effects to the total NME, while the Argonne parametrization lies in between.
Finally, according to our results for the total NMEs in ten candidate nuclei, combined with the observed lower limits on the $0\nu\beta\beta$-decay half-lives, the predicted strongest limits on the effective masses are $|\langle m_\nu\rangle|<0.06~\mathrm{eV}$ for light neutrinos and $|\langle m_{\nu_h}^{-1}\rangle|^{-1}>3.065\times 10^8~\mathrm{GeV}$ for heavy neutrinos.

\begin{acknowledgments}
This work was partially supported by the Major State 973 Program of China (Grant No. 2013CB834400), the National Natural Science Foundation of China (Grants No. 11175002, No. 11335002, No. 11461141002, No. 11575148, No. 11475140, No. 11305134, and No. 11621131001), the Research Fund for the Doctoral Program of Higher Education (Grant No. 20110001110087), the Scientific Discovery through Advanced Computing (SciDAC) program funded by US Department of Energy, Office of Science, Advanced Scientific Computing Research and Nuclear Physics, under Contract No. DE-SC0008641, ER41896, and the DFG cluster of excellence ``Origin and Structure of the Universe'' (www.universe-cluster.de).
\end{acknowledgments}


\begin{thebibliography}{89}%
\makeatletter
\providecommand \@ifxundefined [1]{%
 \@ifx{#1\undefined}
}%
\providecommand \@ifnum [1]{%
 \ifnum #1\expandafter \@firstoftwo
 \else \expandafter \@secondoftwo
 \fi
}%
\providecommand \@ifx [1]{%
 \ifx #1\expandafter \@firstoftwo
 \else \expandafter \@secondoftwo
 \fi
}%
\providecommand \natexlab [1]{#1}%
\providecommand \enquote  [1]{``#1''}%
\providecommand \bibnamefont  [1]{#1}%
\providecommand \bibfnamefont [1]{#1}%
\providecommand \citenamefont [1]{#1}%
\providecommand \href@noop [0]{\@secondoftwo}%
\providecommand \href [0]{\begingroup \@sanitize@url \@href}%
\providecommand \@href[1]{\@@startlink{#1}\@@href}%
\providecommand \@@href[1]{\endgroup#1\@@endlink}%
\providecommand \@sanitize@url [0]{\catcode `\\12\catcode `\$12\catcode
  `\&12\catcode `\#12\catcode `\^12\catcode `\_12\catcode `\%12\relax}%
\providecommand \@@startlink[1]{}%
\providecommand \@@endlink[0]{}%
\providecommand \url  [0]{\begingroup\@sanitize@url \@url }%
\providecommand \@url [1]{\endgroup\@href {#1}{\urlprefix }}%
\providecommand \urlprefix  [0]{URL }%
\providecommand \Eprint [0]{\href }%
\providecommand \doibase [0]{http://dx.doi.org/}%
\providecommand \selectlanguage [0]{\@gobble}%
\providecommand \bibinfo  [0]{\@secondoftwo}%
\providecommand \bibfield  [0]{\@secondoftwo}%
\providecommand \translation [1]{[#1]}%
\providecommand \BibitemOpen [0]{}%
\providecommand \bibitemStop [0]{}%
\providecommand \bibitemNoStop [0]{.\EOS\space}%
\providecommand \EOS [0]{\spacefactor3000\relax}%
\providecommand \BibitemShut  [1]{\csname bibitem#1\endcsname}%
\let\auto@bib@innerbib\@empty
\bibitem [{\citenamefont {Bilenky}(2010)}]{Bilenky2010}%
  \BibitemOpen
  \bibfield  {author} {\bibinfo {author} {\bibfnamefont {S.}~\bibnamefont
  {Bilenky}},\ }\enquote {\bibinfo {title} {Introduction to the physics of
  massive and mixed neutrinos},}\ in\ \href@noop {} {\emph {\bibinfo
  {booktitle} {Lect. Notes Phys.}}},\ Vol.\ \bibinfo {volume} {817}\ (\bibinfo
  {publisher} {Springer, Berlin Heidelberg},\ \bibinfo {year}
  {2010})\BibitemShut {NoStop}%
\bibitem [{\citenamefont {Gando{, \emph{et al.}}}(2016)}]{Gando2016}%
  \BibitemOpen
  \bibfield  {author} {\bibinfo {author} {\bibfnamefont {A.}~\bibnamefont
  {Gando{, \emph{et al.}}}} (\bibinfo {collaboration} {KamLAND-Zen
  Collaboration}),\ }\href {\doibase 10.1103/PhysRevLett.117.082503} {\bibfield
   {journal} {\bibinfo  {journal} {Phys. Rev. Lett.}\ }\textbf {\bibinfo
  {volume} {117}},\ \bibinfo {pages} {082503} (\bibinfo {year}
  {2016})}\BibitemShut {NoStop}%
\bibitem [{\citenamefont {Arnold{, \emph{et al.}}}(2016)}]{Arnold2016}%
  \BibitemOpen
  \bibfield  {author} {\bibinfo {author} {\bibfnamefont {R.}~\bibnamefont
  {Arnold{, \emph{et al.}}}} (\bibinfo {collaboration} {NEMO-3
  Collaboration}),\ }\href {\doibase 10.1103/PhysRevD.94.072003} {\bibfield
  {journal} {\bibinfo  {journal} {Phys. Rev. D}\ }\textbf {\bibinfo {volume}
  {94}},\ \bibinfo {pages} {072003} (\bibinfo {year} {2016})}\BibitemShut
  {NoStop}%
\bibitem [{\citenamefont {Kotila}\ and\ \citenamefont
  {Iachello}(2012)}]{Kotila2012}%
  \BibitemOpen
  \bibfield  {author} {\bibinfo {author} {\bibfnamefont {J.}~\bibnamefont
  {Kotila}}\ and\ \bibinfo {author} {\bibfnamefont {F.}~\bibnamefont
  {Iachello}},\ }\href {\doibase 10.1103/physrevc.85.034316} {\bibfield
  {journal} {\bibinfo  {journal} {Phys. Rev. C}\ }\textbf {\bibinfo {volume}
  {85}},\ \bibinfo {pages} {034316} (\bibinfo {year} {2012})}\BibitemShut
  {NoStop}%
\bibitem [{\citenamefont {Song}\ \emph {et~al.}(2014)\citenamefont {Song},
  \citenamefont {Yao}, \citenamefont {Ring},\ and\ \citenamefont
  {Meng}}]{Song2014}%
  \BibitemOpen
  \bibfield  {author} {\bibinfo {author} {\bibfnamefont {L.~S.}\ \bibnamefont
  {Song}}, \bibinfo {author} {\bibfnamefont {J.~M.}\ \bibnamefont {Yao}},
  \bibinfo {author} {\bibfnamefont {P.}~\bibnamefont {Ring}}, \ and\ \bibinfo
  {author} {\bibfnamefont {J.}~\bibnamefont {Meng}},\ }\href {\doibase
  10.1103/physrevc.90.054309} {\bibfield  {journal} {\bibinfo  {journal} {Phys.
  Rev. C}\ }\textbf {\bibinfo {volume} {90}},\ \bibinfo {pages} {054309}
  (\bibinfo {year} {2014})}\BibitemShut {NoStop}%
\bibitem [{\citenamefont {Yao}\ \emph {et~al.}(2015{\natexlab{a}})\citenamefont
  {Yao}, \citenamefont {Song}, \citenamefont {Hagino}, \citenamefont {Ring},\
  and\ \citenamefont {Meng}}]{Yao2015}%
  \BibitemOpen
  \bibfield  {author} {\bibinfo {author} {\bibfnamefont {J.~M.}\ \bibnamefont
  {Yao}}, \bibinfo {author} {\bibfnamefont {L.~S.}\ \bibnamefont {Song}},
  \bibinfo {author} {\bibfnamefont {K.}~\bibnamefont {Hagino}}, \bibinfo
  {author} {\bibfnamefont {P.}~\bibnamefont {Ring}}, \ and\ \bibinfo {author}
  {\bibfnamefont {J.}~\bibnamefont {Meng}},\ }\href {\doibase
  10.1103/physrevc.91.024316} {\bibfield  {journal} {\bibinfo  {journal} {Phys.
  Rev. C}\ }\textbf {\bibinfo {volume} {91}},\ \bibinfo {pages} {024316}
  (\bibinfo {year} {2015}{\natexlab{a}})}\BibitemShut {NoStop}%
\bibitem [{\citenamefont {Ring}\ \emph {et~al.}(2015)\citenamefont {Ring},
  \citenamefont {Yao}, \citenamefont {Song}, \citenamefont {Hagino},\ and\
  \citenamefont {Meng}}]{Ring2015}%
  \BibitemOpen
  \bibfield  {author} {\bibinfo {author} {\bibfnamefont {P.}~\bibnamefont
  {Ring}}, \bibinfo {author} {\bibfnamefont {J.~M.}\ \bibnamefont {Yao}},
  \bibinfo {author} {\bibfnamefont {L.~S.}\ \bibnamefont {Song}}, \bibinfo
  {author} {\bibfnamefont {K.}~\bibnamefont {Hagino}}, \ and\ \bibinfo {author}
  {\bibfnamefont {J.}~\bibnamefont {Meng}},\ }\href {\doibase
  http://dx.doi.org/10.1063/1.4932283} {\bibfield  {journal} {\bibinfo
  {journal} {AIP Conf. Proc.}\ }\textbf {\bibinfo {volume} {1681}},\ \bibinfo
  {pages} {050008} (\bibinfo {year} {2015})}\BibitemShut {NoStop}%
\bibitem [{\citenamefont {Yao}\ and\ \citenamefont {Engel}(2016)}]{Yao2016}%
  \BibitemOpen
  \bibfield  {author} {\bibinfo {author} {\bibfnamefont {J.~M.}\ \bibnamefont
  {Yao}}\ and\ \bibinfo {author} {\bibfnamefont {J.}~\bibnamefont {Engel}},\
  }\href {\doibase 10.1103/PhysRevC.94.014306} {\bibfield  {journal} {\bibinfo
  {journal} {Phys. Rev. C}\ }\textbf {\bibinfo {volume} {94}},\ \bibinfo
  {pages} {014306} (\bibinfo {year} {2016})}\BibitemShut {NoStop}%
\bibitem {Meng2016}%
  \BibitemOpen
  \href@noop {} {}\bibinfo {title} {\emph{Relativistic Density Functional for Nuclear Structure}}, \ \bibinfo {howpublished} {International Review of Nuclear Physics, edited by J. Meng, Vol. 10 (World Scientific, Singapore, 2016)}\BibitemShut {NoStop}%
\bibitem [{\citenamefont {Haxton}\ and\ \citenamefont
  {Stephenson}(1984)}]{Haxton1984}%
  \BibitemOpen
  \bibfield  {author} {\bibinfo {author} {\bibfnamefont {W.}~\bibnamefont
  {Haxton}}\ and\ \bibinfo {author} {\bibfnamefont {G.}~\bibnamefont
  {Stephenson}},\ }\href@noop {} {\bibfield  {journal} {\bibinfo  {journal}
  {Prog. Part. Nucl. Phys.}\ }\textbf {\bibinfo {volume} {12}},\ \bibinfo
  {pages} {409} (\bibinfo {year} {1984})}\BibitemShut {NoStop}%
\bibitem [{\citenamefont {Wu}\ \emph {et~al.}(1985)\citenamefont {Wu},
  \citenamefont {Song}, \citenamefont {Kuo}, \citenamefont {Cheng},\ and\
  \citenamefont {Strottman}}]{Wu1985}%
  \BibitemOpen
  \bibfield  {author} {\bibinfo {author} {\bibfnamefont {H.}~\bibnamefont
  {Wu}}, \bibinfo {author} {\bibfnamefont {H.}~\bibnamefont {Song}}, \bibinfo
  {author} {\bibfnamefont {T.}~\bibnamefont {Kuo}}, \bibinfo {author}
  {\bibfnamefont {W.}~\bibnamefont {Cheng}}, \ and\ \bibinfo {author}
  {\bibfnamefont {D.}~\bibnamefont {Strottman}},\ }\href {\doibase
  http://dx.doi.org/10.1016/0370-2693(85)90911-6} {\bibfield  {journal}
  {\bibinfo  {journal} {Phys. Lett. B}\ }\textbf {\bibinfo {volume} {162}},\
  \bibinfo {pages} {227} (\bibinfo {year} {1985})}\BibitemShut {NoStop}%
\bibitem [{\citenamefont {Retamosa}\ \emph {et~al.}(1995)\citenamefont
  {Retamosa}, \citenamefont {Caurier},\ and\ \citenamefont
  {Nowacki}}]{Retamosa1995}%
  \BibitemOpen
  \bibfield  {author} {\bibinfo {author} {\bibfnamefont {J.}~\bibnamefont
  {Retamosa}}, \bibinfo {author} {\bibfnamefont {E.}~\bibnamefont {Caurier}}, \
  and\ \bibinfo {author} {\bibfnamefont {F.}~\bibnamefont {Nowacki}},\
  }\href@noop {} {\bibfield  {journal} {\bibinfo  {journal} {Phys. Rev. C}\
  }\textbf {\bibinfo {volume} {51}},\ \bibinfo {pages} {371} (\bibinfo {year}
  {1995})}\BibitemShut {NoStop}%
\bibitem [{\citenamefont {Caurier}\ \emph {et~al.}(1996)\citenamefont
  {Caurier}, \citenamefont {Nowacki}, \citenamefont {Poves},\ and\
  \citenamefont {Retamosa}}]{Caurier1996}%
  \BibitemOpen
  \bibfield  {author} {\bibinfo {author} {\bibfnamefont {E.}~\bibnamefont
  {Caurier}}, \bibinfo {author} {\bibfnamefont {F.}~\bibnamefont {Nowacki}},
  \bibinfo {author} {\bibfnamefont {A.}~\bibnamefont {Poves}}, \ and\ \bibinfo
  {author} {\bibfnamefont {J.}~\bibnamefont {Retamosa}},\ }\href@noop {}
  {\bibfield  {journal} {\bibinfo  {journal} {Phys. Rev. lett.}\ }\textbf
  {\bibinfo {volume} {77}},\ \bibinfo {pages} {1954} (\bibinfo {year}
  {1996})}\BibitemShut {NoStop}%
\bibitem [{\citenamefont {Caurier}\ \emph {et~al.}(2008)\citenamefont
  {Caurier}, \citenamefont {Nowacki},\ and\ \citenamefont
  {Poves}}]{Caurier2008}%
  \BibitemOpen
  \bibfield  {author} {\bibinfo {author} {\bibfnamefont {E.}~\bibnamefont
  {Caurier}}, \bibinfo {author} {\bibfnamefont {F.}~\bibnamefont {Nowacki}}, \
  and\ \bibinfo {author} {\bibfnamefont {A.}~\bibnamefont {Poves}},\ }\href
  {\doibase 10.1140/epja/i2007-10527-x} {\bibfield  {journal} {\bibinfo
  {journal} {Eur. Phys. J. A}\ }\textbf {\bibinfo {volume} {36}},\ \bibinfo
  {pages} {195} (\bibinfo {year} {2008})}\BibitemShut {NoStop}%
\bibitem [{\citenamefont {Men{\'{e}}ndez}\ \emph
  {et~al.}(2009{\natexlab{a}})\citenamefont {Men{\'{e}}ndez}, \citenamefont
  {Poves}, \citenamefont {Caurier},\ and\ \citenamefont
  {Nowacki}}]{Menendez2009}%
  \BibitemOpen
  \bibfield  {author} {\bibinfo {author} {\bibfnamefont {J.}~\bibnamefont
  {Men{\'{e}}ndez}}, \bibinfo {author} {\bibfnamefont {A.}~\bibnamefont
  {Poves}}, \bibinfo {author} {\bibfnamefont {E.}~\bibnamefont {Caurier}}, \
  and\ \bibinfo {author} {\bibfnamefont {F.}~\bibnamefont {Nowacki}},\ }\href
  {\doibase 10.1016/j.nuclphysa.2008.12.005} {\bibfield  {journal} {\bibinfo
  {journal} {Nucl. Phys. A}\ }\textbf {\bibinfo {volume} {818}},\ \bibinfo
  {pages} {139} (\bibinfo {year} {2009}{\natexlab{a}})}\BibitemShut {NoStop}%
\bibitem [{\citenamefont {Men{\'{e}}ndez}\ \emph
  {et~al.}(2009{\natexlab{b}})\citenamefont {Men{\'{e}}ndez}, \citenamefont
  {Poves}, \citenamefont {Caurier},\ and\ \citenamefont
  {Nowacki}}]{Menendez2009a}%
  \BibitemOpen
  \bibfield  {author} {\bibinfo {author} {\bibfnamefont {J.}~\bibnamefont
  {Men{\'{e}}ndez}}, \bibinfo {author} {\bibfnamefont {A.}~\bibnamefont
  {Poves}}, \bibinfo {author} {\bibfnamefont {E.}~\bibnamefont {Caurier}}, \
  and\ \bibinfo {author} {\bibfnamefont {F.}~\bibnamefont {Nowacki}},\ }\href
  {\doibase 10.1103/physrevc.80.048501} {\bibfield  {journal} {\bibinfo
  {journal} {Phys. Rev. C}\ }\textbf {\bibinfo {volume} {80}},\ \bibinfo
  {pages} {048501} (\bibinfo {year} {2009}{\natexlab{b}})}\BibitemShut
  {NoStop}%
\bibitem [{\citenamefont {Men{\'{e}}ndez}\ \emph {et~al.}(2011)\citenamefont
  {Men{\'{e}}ndez}, \citenamefont {Gazit},\ and\ \citenamefont
  {Schwenk}}]{Menendez2011}%
  \BibitemOpen
  \bibfield  {author} {\bibinfo {author} {\bibfnamefont {J.}~\bibnamefont
  {Men{\'{e}}ndez}}, \bibinfo {author} {\bibfnamefont {D.}~\bibnamefont
  {Gazit}}, \ and\ \bibinfo {author} {\bibfnamefont {A.}~\bibnamefont
  {Schwenk}},\ }\href {\doibase 10.1103/physrevlett.107.062501} {\bibfield
  {journal} {\bibinfo  {journal} {Phys. Rev. Lett.}\ }\textbf {\bibinfo
  {volume} {107}},\ \bibinfo {pages} {062501} (\bibinfo {year}
  {2011})}\BibitemShut {NoStop}%
\bibitem [{\citenamefont {Sen'kov}\ and\ \citenamefont
  {Horoi}(2013)}]{Senkov2013}%
  \BibitemOpen
  \bibfield  {author} {\bibinfo {author} {\bibfnamefont {R.~A.}\ \bibnamefont
  {Sen'kov}}\ and\ \bibinfo {author} {\bibfnamefont {M.}~\bibnamefont
  {Horoi}},\ }\href {\doibase 10.1103/PhysRevC.88.064312} {\bibfield  {journal}
  {\bibinfo  {journal} {Phys. Rev. C}\ }\textbf {\bibinfo {volume} {88}},\
  \bibinfo {pages} {064312} (\bibinfo {year} {2013})}\BibitemShut {NoStop}%
\bibitem [{\citenamefont {Sen'kov}\ \emph {et~al.}(2014)\citenamefont
  {Sen'kov}, \citenamefont {Horoi},\ and\ \citenamefont {Brown}}]{Senkov2014}%
  \BibitemOpen
  \bibfield  {author} {\bibinfo {author} {\bibfnamefont {R.~A.}\ \bibnamefont
  {Sen'kov}}, \bibinfo {author} {\bibfnamefont {M.}~\bibnamefont {Horoi}}, \
  and\ \bibinfo {author} {\bibfnamefont {B.~A.}\ \bibnamefont {Brown}},\ }\href
  {\doibase 10.1103/physrevc.89.054304} {\bibfield  {journal} {\bibinfo
  {journal} {Phys. Rev. C}\ }\textbf {\bibinfo {volume} {89}},\ \bibinfo
  {pages} {054304} (\bibinfo {year} {2014})}\BibitemShut {NoStop}%
\bibitem [{\citenamefont {Sen'kov}\ and\ \citenamefont
  {Horoi}(2014)}]{Senkov2014a}%
  \BibitemOpen
  \bibfield  {author} {\bibinfo {author} {\bibfnamefont {R.~A.}\ \bibnamefont
  {Sen'kov}}\ and\ \bibinfo {author} {\bibfnamefont {M.}~\bibnamefont
  {Horoi}},\ }\href {\doibase 10.1103/physrevc.90.051301} {\bibfield  {journal}
  {\bibinfo  {journal} {Phys. Rev. C}\ }\textbf {\bibinfo {volume} {90}},\ \bibinfo
  {pages} {051301} (\bibinfo {year} {2014})}\BibitemShut {NoStop}%
\bibitem [{\citenamefont {Horoi}\ and\ \citenamefont
  {Stoica}(2010)}]{Horoi2010}%
  \BibitemOpen
  \bibfield  {author} {\bibinfo {author} {\bibfnamefont {M.}~\bibnamefont
  {Horoi}}\ and\ \bibinfo {author} {\bibfnamefont {S.}~\bibnamefont {Stoica}},\
  }\href {\doibase 10.1103/physrevc.81.024321} {\bibfield  {journal} {\bibinfo
  {journal} {Phys. Rev. C}\ }\textbf {\bibinfo {volume} {81}},\ \bibinfo
  {pages} {024321} (\bibinfo {year} {2010})}\BibitemShut {NoStop}%
\bibitem [{\citenamefont {Horoi}(2013)}]{Horoi2013}%
  \BibitemOpen
  \bibfield  {author} {\bibinfo {author} {\bibfnamefont {M.}~\bibnamefont
  {Horoi}},\ }\href {\doibase 10.1103/physrevc.87.014320} {\bibfield  {journal}
  {\bibinfo  {journal} {Phys. Rev. C}\ }\textbf {\bibinfo {volume} {87}},\
  \bibinfo {pages} {014320} (\bibinfo {year} {2013})}\BibitemShut {NoStop}%
\bibitem [{\citenamefont {Neacsu}\ and\ \citenamefont
  {Horoi}(2015)}]{Neacsu2015}%
  \BibitemOpen
  \bibfield  {author} {\bibinfo {author} {\bibfnamefont {A.}~\bibnamefont
  {Neacsu}}\ and\ \bibinfo {author} {\bibfnamefont {M.}~\bibnamefont {Horoi}},\
  }\href {\doibase 10.1103/physrevc.91.024309} {\bibfield  {journal} {\bibinfo
  {journal} {Phys. Rev. C}\ }\textbf {\bibinfo {volume} {91}},\ \bibinfo
  {pages} {024309} (\bibinfo {year} {2015})}\BibitemShut {NoStop}%
\bibitem [{\citenamefont {Horoi}\ and\ \citenamefont
  {Neacsu}(2016)}]{Horoi2016}%
  \BibitemOpen
  \bibfield  {author} {\bibinfo {author} {\bibfnamefont {M.}~\bibnamefont
  {Horoi}}\ and\ \bibinfo {author} {\bibfnamefont {A.}~\bibnamefont {Neacsu}},\
  }\href {\doibase 10.1103/physrevc.93.024308} {\bibfield  {journal} {\bibinfo
  {journal} {Phys. Rev. C}\ }\textbf {\bibinfo {volume} {93}},\ \bibinfo
  {pages} {024308} (\bibinfo {year} {2016})}\BibitemShut {NoStop}%
\bibitem [{\citenamefont {Iwata}\ \emph {et~al.}(2016)\citenamefont {Iwata},
  \citenamefont {Shimizu}, \citenamefont {Otsuka}, \citenamefont {Utsuno},
  \citenamefont {Men{\'{e}}ndez}, \citenamefont {Honma},\ and\ \citenamefont
  {Abe}}]{Iwata2016}%
  \BibitemOpen
  \bibfield  {author} {\bibinfo {author} {\bibfnamefont {Y.}~\bibnamefont
  {Iwata}}, \bibinfo {author} {\bibfnamefont {N.}~\bibnamefont {Shimizu}},
  \bibinfo {author} {\bibfnamefont {T.}~\bibnamefont {Otsuka}}, \bibinfo
  {author} {\bibfnamefont {Y.}~\bibnamefont {Utsuno}}, \bibinfo {author}
  {\bibfnamefont {J.}~\bibnamefont {Men{\'{e}}ndez}}, \bibinfo {author}
  {\bibfnamefont {M.}~\bibnamefont {Honma}}, \ and\ \bibinfo {author}
  {\bibfnamefont {T.}~\bibnamefont {Abe}},\ }\href {\doibase
  10.1103/physrevlett.116.112502} {\bibfield  {journal} {\bibinfo  {journal}
  {Phys. Rev. Lett.}\ }\textbf {\bibinfo {volume} {116}},\ \bibinfo {pages}
  {112502} (\bibinfo {year} {2016})}\BibitemShut {NoStop}%
\bibitem [{\citenamefont {Tomoda}\ and\ \citenamefont
  {Faessler}(1987)}]{Tomoda1987}%
  \BibitemOpen
  \bibfield  {author} {\bibinfo {author} {\bibfnamefont {T.}~\bibnamefont
  {Tomoda}}\ and\ \bibinfo {author} {\bibfnamefont {A.}~\bibnamefont
  {Faessler}},\ }\href@noop {} {\bibfield  {journal} {\bibinfo  {journal}
  {Phys. Lett. B}\ }\textbf {\bibinfo {volume} {199}},\ \bibinfo {pages} {475}
  (\bibinfo {year} {1987})}\BibitemShut {NoStop}%
\bibitem [{\citenamefont {Muto}\ \emph {et~al.}(1989)\citenamefont {Muto},
  \citenamefont {Bender},\ and\ \citenamefont {Klapdor}}]{Muto1989}%
  \BibitemOpen
  \bibfield  {author} {\bibinfo {author} {\bibfnamefont {K.}~\bibnamefont
  {Muto}}, \bibinfo {author} {\bibfnamefont {E.}~\bibnamefont {Bender}}, \ and\
  \bibinfo {author} {\bibfnamefont {H.}~\bibnamefont {Klapdor}},\ }\href@noop
  {} {\bibfield  {journal} {\bibinfo  {journal} {Z. Phy. A}\ }\textbf {\bibinfo
  {volume} {334}},\ \bibinfo {pages} {187} (\bibinfo {year}
  {1989})}\BibitemShut {NoStop}%
\bibitem [{\citenamefont {Staudt}\ \emph {et~al.}(1990)\citenamefont {Staudt},
  \citenamefont {Kuo},\ and\ \citenamefont
  {Klapdor-Kleingrothaus}}]{Staudt1990}%
  \BibitemOpen
  \bibfield  {author} {\bibinfo {author} {\bibfnamefont {A.}~\bibnamefont
  {Staudt}}, \bibinfo {author} {\bibfnamefont {T.}~\bibnamefont {Kuo}}, \ and\
  \bibinfo {author} {\bibfnamefont {H.}~\bibnamefont {Klapdor-Kleingrothaus}},\
  }\href@noop {} {\bibfield  {journal} {\bibinfo  {journal} {Phys. Lett. B}\
  }\textbf {\bibinfo {volume} {242}},\ \bibinfo {pages} {17} (\bibinfo {year}
  {1990})}\BibitemShut {NoStop}%
\bibitem [{\citenamefont {Stout}\ and\ \citenamefont {Kuo}(1992)}]{Stout1992}%
  \BibitemOpen
  \bibfield  {author} {\bibinfo {author} {\bibfnamefont {D.}~\bibnamefont
  {Stout}}\ and\ \bibinfo {author} {\bibfnamefont {T.}~\bibnamefont {Kuo}},\
  }\href@noop {} {\bibfield  {journal} {\bibinfo  {journal} {Phys. Rev. Lett.}\
  }\textbf {\bibinfo {volume} {69}},\ \bibinfo {pages} {1900} (\bibinfo {year}
  {1992})}\BibitemShut {NoStop}%
\bibitem [{\citenamefont {Staudt}\ \emph {et~al.}(1992)\citenamefont {Staudt},
  \citenamefont {Kuo},\ and\ \citenamefont
  {Klapdor-Kleingrothaus}}]{Staudt1992}%
  \BibitemOpen
  \bibfield  {author} {\bibinfo {author} {\bibfnamefont {A.}~\bibnamefont
  {Staudt}}, \bibinfo {author} {\bibfnamefont {T.}~\bibnamefont {Kuo}}, \ and\
  \bibinfo {author} {\bibfnamefont {H.}~\bibnamefont {Klapdor-Kleingrothaus}},\
  }\href@noop {} {\bibfield  {journal} {\bibinfo  {journal} {Phys. Rev. C}\
  }\textbf {\bibinfo {volume} {46}},\ \bibinfo {pages} {871} (\bibinfo {year}
  {1992})}\BibitemShut {NoStop}%
\bibitem [{\citenamefont {Pantis}\ \emph {et~al.}(1992)\citenamefont {Pantis},
  \citenamefont {Faessler}, \citenamefont {Kaminski},\ and\ \citenamefont
  {Vergados}}]{Pantis1992}%
  \BibitemOpen
  \bibfield  {author} {\bibinfo {author} {\bibfnamefont {G.}~\bibnamefont
  {Pantis}}, \bibinfo {author} {\bibfnamefont {A.}~\bibnamefont {Faessler}},
  \bibinfo {author} {\bibfnamefont {W.}~\bibnamefont {Kaminski}}, \ and\
  \bibinfo {author} {\bibfnamefont {J.}~\bibnamefont {Vergados}},\ }\href@noop
  {} {\bibfield  {journal} {\bibinfo  {journal} {J. Phys. G: Nucl. Part.
  Phys.}\ }\textbf {\bibinfo {volume} {18}},\ \bibinfo {pages} {605} (\bibinfo
  {year} {1992})}\BibitemShut {NoStop}%
\bibitem [{\citenamefont {Kortelainen}\ and\ \citenamefont
  {Suhonen}(2007{\natexlab{a}})}]{Kortelainen2007}%
  \BibitemOpen
  \bibfield  {author} {\bibinfo {author} {\bibfnamefont {M.}~\bibnamefont
  {Kortelainen}}\ and\ \bibinfo {author} {\bibfnamefont {J.}~\bibnamefont
  {Suhonen}},\ }\href {\doibase 10.1103/physrevc.76.024315} {\bibfield
  {journal} {\bibinfo  {journal} {Phys. Rev. C}\ }\textbf {\bibinfo {volume}
  {76}},\ \bibinfo {pages} {024315} (\bibinfo {year}
  {2007}{\natexlab{a}})}\BibitemShut {NoStop}%
\bibitem [{\citenamefont {Kortelainen}\ and\ \citenamefont
  {Suhonen}(2007{\natexlab{b}})}]{Kortelainen2007a}%
  \BibitemOpen
  \bibfield  {author} {\bibinfo {author} {\bibfnamefont {M.}~\bibnamefont
  {Kortelainen}}\ and\ \bibinfo {author} {\bibfnamefont {J.}~\bibnamefont
  {Suhonen}},\ }\href {\doibase 10.1103/physrevc.75.051303} {\bibfield
  {journal} {\bibinfo  {journal} {Phys. Rev. C}\ }\textbf {\bibinfo {volume}
  {75}},\ \bibinfo {pages} {051303} (\bibinfo {year}
  {2007}{\natexlab{b}})}\BibitemShut {NoStop}%
\bibitem [{\citenamefont {Pantis}\ \emph {et~al.}(1996)\citenamefont {Pantis},
  \citenamefont {{\v{S}}imkovic}, \citenamefont {Vergados},\ and\ \citenamefont
  {Faessler}}]{Pantis1996}%
  \BibitemOpen
  \bibfield  {author} {\bibinfo {author} {\bibfnamefont {G.}~\bibnamefont
  {Pantis}}, \bibinfo {author} {\bibfnamefont {F.}~\bibnamefont
  {{\v{S}}imkovic}}, \bibinfo {author} {\bibfnamefont {J.}~\bibnamefont
  {Vergados}}, \ and\ \bibinfo {author} {\bibfnamefont {A.}~\bibnamefont
  {Faessler}},\ }\href@noop {} {\bibfield  {journal} {\bibinfo  {journal}
  {Phys. Rev. C}\ }\textbf {\bibinfo {volume} {53}},\ \bibinfo {pages} {695}
  (\bibinfo {year} {1996})}\BibitemShut {NoStop}%
\bibitem [{\citenamefont {{\v{S}}imkovic}\ \emph {et~al.}(1997)\citenamefont
  {{\v{S}}imkovic}, \citenamefont {Schwieger}, \citenamefont {Veselsk{\`y}},
  \citenamefont {Pantis},\ and\ \citenamefont {Faessler}}]{Simkovic1997}%
  \BibitemOpen
  \bibfield  {author} {\bibinfo {author} {\bibfnamefont {F.}~\bibnamefont
  {{\v{S}}imkovic}}, \bibinfo {author} {\bibfnamefont {J.}~\bibnamefont
  {Schwieger}}, \bibinfo {author} {\bibfnamefont {M.}~\bibnamefont
  {Veselsk{\`y}}}, \bibinfo {author} {\bibfnamefont {G.}~\bibnamefont
  {Pantis}}, \ and\ \bibinfo {author} {\bibfnamefont {A.}~\bibnamefont
  {Faessler}},\ }\href@noop {} {\bibfield  {journal} {\bibinfo  {journal}
  {Phys. Lett. B}\ }\textbf {\bibinfo {volume} {393}},\ \bibinfo {pages} {267}
  (\bibinfo {year} {1997})}\BibitemShut {NoStop}%
\bibitem [{\citenamefont {{\v{S}}imkovic}\ \emph {et~al.}(1999)\citenamefont
  {{\v{S}}imkovic}, \citenamefont {Pantis}, \citenamefont {Vergados},\ and\
  \citenamefont {Faessler}}]{Simkovic1999}%
  \BibitemOpen
  \bibfield  {author} {\bibinfo {author} {\bibfnamefont {F.}~\bibnamefont
  {{\v{S}}imkovic}}, \bibinfo {author} {\bibfnamefont {G.}~\bibnamefont
  {Pantis}}, \bibinfo {author} {\bibfnamefont {J.}~\bibnamefont {Vergados}}, \
  and\ \bibinfo {author} {\bibfnamefont {A.}~\bibnamefont {Faessler}},\
  }\href@noop {} {\bibfield  {journal} {\bibinfo  {journal} {Phys. Rev. C}\
  }\textbf {\bibinfo {volume} {60}},\ \bibinfo {pages} {055502} (\bibinfo
  {year} {1999})}\BibitemShut {NoStop}%
\bibitem [{\citenamefont {Rodin}\ \emph {et~al.}(2006)\citenamefont {Rodin},
  \citenamefont {Faessler}, \citenamefont {{\v{S}}imkovic},\ and\ \citenamefont
  {Vogel}}]{Rodin2006}%
  \BibitemOpen
  \bibfield  {author} {\bibinfo {author} {\bibfnamefont {V.}~\bibnamefont
  {Rodin}}, \bibinfo {author} {\bibfnamefont {A.}~\bibnamefont {Faessler}},
  \bibinfo {author} {\bibfnamefont {F.}~\bibnamefont {{\v{S}}imkovic}}, \ and\
  \bibinfo {author} {\bibfnamefont {P.}~\bibnamefont {Vogel}},\ }\href
  {\doibase 10.1016/j.nuclphysa.2005.12.004} {\bibfield  {journal} {\bibinfo
  {journal} {Nucl. Phys. A}\ }\textbf {\bibinfo {volume} {766}},\ \bibinfo
  {pages} {107} (\bibinfo {year} {2006})}\BibitemShut {NoStop}%
\bibitem [{\citenamefont {Rodin}\ \emph {et~al.}(2007)\citenamefont {Rodin},
  \citenamefont {Faessler}, \citenamefont {{\v{S}}imkovic},\ and\ \citenamefont
  {Vogel}}]{Rodin2007}%
  \BibitemOpen
  \bibfield  {author} {\bibinfo {author} {\bibfnamefont {V.}~\bibnamefont
  {Rodin}}, \bibinfo {author} {\bibfnamefont {A.}~\bibnamefont {Faessler}},
  \bibinfo {author} {\bibfnamefont {F.}~\bibnamefont {{\v{S}}imkovic}}, \ and\
  \bibinfo {author} {\bibfnamefont {P.}~\bibnamefont {Vogel}},\ }\href
  {\doibase http://dx.doi.org/10.1016/j.nuclphysa.2007.06.014} {\bibfield
  {journal} {\bibinfo  {journal} {Nucl. Phys. A}\ }\textbf {\bibinfo {volume}
  {793}},\ \bibinfo {pages} {213} (\bibinfo {year} {2007})}\BibitemShut
  {NoStop}%
\bibitem [{\citenamefont {{\v{S}}imkovic}\ \emph {et~al.}(2008)\citenamefont
  {{\v{S}}imkovic}, \citenamefont {Faessler}, \citenamefont {Rodin},
  \citenamefont {Vogel},\ and\ \citenamefont {Engel}}]{Simkovic2008}%
  \BibitemOpen
  \bibfield  {author} {\bibinfo {author} {\bibfnamefont {F.}~\bibnamefont
  {{\v{S}}imkovic}}, \bibinfo {author} {\bibfnamefont {A.}~\bibnamefont
  {Faessler}}, \bibinfo {author} {\bibfnamefont {V.}~\bibnamefont {Rodin}},
  \bibinfo {author} {\bibfnamefont {P.}~\bibnamefont {Vogel}}, \ and\ \bibinfo
  {author} {\bibfnamefont {J.}~\bibnamefont {Engel}},\ }\href {\doibase
  10.1103/physrevc.77.045503} {\bibfield  {journal} {\bibinfo  {journal} {Phys.
  Rev. C}\ }\textbf {\bibinfo {volume} {77}},\ \bibinfo {pages} {045503}
  (\bibinfo {year} {2008})}\BibitemShut {NoStop}%
\bibitem [{\citenamefont {{\v{S}}imkovic}\ \emph
  {et~al.}(2009{\natexlab{a}})\citenamefont {{\v{S}}imkovic}, \citenamefont
  {Faessler},\ and\ \citenamefont {Vogel}}]{Simkovic2009}%
  \BibitemOpen
  \bibfield  {author} {\bibinfo {author} {\bibfnamefont {F.}~\bibnamefont
  {{\v{S}}imkovic}}, \bibinfo {author} {\bibfnamefont {A.}~\bibnamefont
  {Faessler}}, \ and\ \bibinfo {author} {\bibfnamefont {P.}~\bibnamefont
  {Vogel}},\ }\href {\doibase 10.1103/physrevc.79.015502} {\bibfield  {journal}
  {\bibinfo  {journal} {Phys. Rev. C}\ }\textbf {\bibinfo {volume} {79}},\
  \bibinfo {pages} {015502} (\bibinfo {year} {2009}{\natexlab{a}})}\BibitemShut
  {NoStop}%
\bibitem [{\citenamefont {Fang}\ \emph {et~al.}(2010)\citenamefont {Fang},
  \citenamefont {Faessler}, \citenamefont {Rodin},\ and\ \citenamefont
  {{\v{S}}imkovic}}]{Fang2010}%
  \BibitemOpen
  \bibfield  {author} {\bibinfo {author} {\bibfnamefont {D.-L.}\ \bibnamefont
  {Fang}}, \bibinfo {author} {\bibfnamefont {A.}~\bibnamefont {Faessler}},
  \bibinfo {author} {\bibfnamefont {V.}~\bibnamefont {Rodin}}, \ and\ \bibinfo
  {author} {\bibfnamefont {F.}~\bibnamefont {{\v{S}}imkovic}},\ }\href
  {\doibase 10.1103/physrevc.82.051301} {\bibfield  {journal} {\bibinfo
  {journal} {Phys. Rev. C}\ }\textbf {\bibinfo {volume} {82}},\ \bibinfo
  {pages} {051301} (\bibinfo {year} {2010})}\BibitemShut {NoStop}%
\bibitem [{\citenamefont {Fang}\ \emph {et~al.}(2011)\citenamefont {Fang},
  \citenamefont {Faessler}, \citenamefont {Rodin},\ and\ \citenamefont
  {{\v{S}}imkovic}}]{Fang2011}%
  \BibitemOpen
  \bibfield  {author} {\bibinfo {author} {\bibfnamefont {D.-L.}\ \bibnamefont
  {Fang}}, \bibinfo {author} {\bibfnamefont {A.}~\bibnamefont {Faessler}},
  \bibinfo {author} {\bibfnamefont {V.}~\bibnamefont {Rodin}}, \ and\ \bibinfo
  {author} {\bibfnamefont {F.}~\bibnamefont {{\v{S}}imkovic}},\ }\href
  {\doibase 10.1103/physrevc.83.034320} {\bibfield  {journal} {\bibinfo
  {journal} {Phys. Rev. C}\ }\textbf {\bibinfo {volume} {83}},\ \bibinfo
  {pages} {034320} (\bibinfo {year} {2011})}\BibitemShut {NoStop}%
\bibitem [{\citenamefont {Mustonen}\ and\ \citenamefont
  {Engel}(2013)}]{Mustonen2013}%
  \BibitemOpen
  \bibfield  {author} {\bibinfo {author} {\bibfnamefont {M.~T.}\ \bibnamefont
  {Mustonen}}\ and\ \bibinfo {author} {\bibfnamefont {J.}~\bibnamefont
  {Engel}},\ }\href {\doibase 10.1103/physrevc.87.064302} {\bibfield  {journal}
  {\bibinfo  {journal} {Phys. Rev. C}\ }\textbf {\bibinfo {volume} {87}},\
  \bibinfo {pages} {064302} (\bibinfo {year} {2013})}\BibitemShut {NoStop}%
\bibitem [{\citenamefont {Terasaki}(2015)}]{Terasaki2015}%
  \BibitemOpen
  \bibfield  {author} {\bibinfo {author} {\bibfnamefont {J.}~\bibnamefont
  {Terasaki}},\ }\href {\doibase 10.1103/physrevc.91.034318} {\bibfield
  {journal} {\bibinfo  {journal} {Phys. Rev. C}\ }\textbf {\bibinfo {volume}
  {91}},\ \bibinfo {pages} {034318} (\bibinfo {year} {2015})}\BibitemShut
  {NoStop}%
\bibitem [{\citenamefont {{\v{S}}imkovic}\ \emph {et~al.}(2013)\citenamefont
  {{\v{S}}imkovic}, \citenamefont {Rodin}, \citenamefont {Faessler},\ and\
  \citenamefont {Vogel}}]{Simkovic2013}%
  \BibitemOpen
  \bibfield  {author} {\bibinfo {author} {\bibfnamefont {F.}~\bibnamefont
  {{\v{S}}imkovic}}, \bibinfo {author} {\bibfnamefont {V.}~\bibnamefont
  {Rodin}}, \bibinfo {author} {\bibfnamefont {A.}~\bibnamefont {Faessler}}, \
  and\ \bibinfo {author} {\bibfnamefont {P.}~\bibnamefont {Vogel}},\ }\href
  {\doibase 10.1103/physrevc.87.045501} {\bibfield  {journal} {\bibinfo
  {journal} {Phys. Rev. C}\ }\textbf {\bibinfo {volume} {87}},\ \bibinfo
  {pages} {045501} (\bibinfo {year} {2013})}\BibitemShut {NoStop}%
\bibitem [{\citenamefont {Fang}\ \emph {et~al.}(2015)\citenamefont {Fang},
  \citenamefont {Faessler},\ and\ \citenamefont {{\v{S}}imkovic}}]{Fang2015}%
  \BibitemOpen
  \bibfield  {author} {\bibinfo {author} {\bibfnamefont {D.-L.}\ \bibnamefont
  {Fang}}, \bibinfo {author} {\bibfnamefont {A.}~\bibnamefont {Faessler}}, \
  and\ \bibinfo {author} {\bibfnamefont {F.}~\bibnamefont {{\v{S}}imkovic}},\
  }\href {\doibase 10.1103/physrevc.92.044301} {\bibfield  {journal} {\bibinfo
  {journal} {Phys. Rev. C}\ }\textbf {\bibinfo {volume} {92}},\ \bibinfo
  {pages} {044301} (\bibinfo {year} {2015})}\BibitemShut {NoStop}%
\bibitem [{\citenamefont {Chaturvedi}\ \emph {et~al.}(2008)\citenamefont
  {Chaturvedi}, \citenamefont {Chandra}, \citenamefont {Rath}, \citenamefont
  {Raina},\ and\ \citenamefont {Hirsch}}]{Chaturvedi2008}%
  \BibitemOpen
  \bibfield  {author} {\bibinfo {author} {\bibfnamefont {K.}~\bibnamefont
  {Chaturvedi}}, \bibinfo {author} {\bibfnamefont {R.}~\bibnamefont {Chandra}},
  \bibinfo {author} {\bibfnamefont {P.~K.}\ \bibnamefont {Rath}}, \bibinfo
  {author} {\bibfnamefont {P.~K.}\ \bibnamefont {Raina}}, \ and\ \bibinfo
  {author} {\bibfnamefont {J.~G.}\ \bibnamefont {Hirsch}},\ }\href {\doibase
  10.1103/physrevc.78.054302} {\bibfield  {journal} {\bibinfo  {journal} {Phys.
  Rev. C}\ }\textbf {\bibinfo {volume} {78}},\ \bibinfo {pages} {054302}
  (\bibinfo {year} {2008})}\BibitemShut {NoStop}%
\bibitem [{\citenamefont {Chandra}\ \emph {et~al.}(2009)\citenamefont
  {Chandra}, \citenamefont {Chaturvedi}, \citenamefont {Rath}, \citenamefont
  {Raina},\ and\ \citenamefont {Hirsch}}]{Chandra2009}%
  \BibitemOpen
  \bibfield  {author} {\bibinfo {author} {\bibfnamefont {R.}~\bibnamefont
  {Chandra}}, \bibinfo {author} {\bibfnamefont {K.}~\bibnamefont {Chaturvedi}},
  \bibinfo {author} {\bibfnamefont {P.}~\bibnamefont {Rath}}, \bibinfo {author}
  {\bibfnamefont {P.}~\bibnamefont {Raina}}, \ and\ \bibinfo {author}
  {\bibfnamefont {J.}~\bibnamefont {Hirsch}},\ }\href@noop {} {\bibfield
  {journal} {\bibinfo  {journal} {Europhys. Lett.}\ }\textbf {\bibinfo {volume}
  {86}},\ \bibinfo {pages} {32001} (\bibinfo {year} {2009})}\BibitemShut
  {NoStop}%
\bibitem [{\citenamefont {Rath}\ \emph {et~al.}(2009)\citenamefont {Rath},
  \citenamefont {Chandra}, \citenamefont {Chaturvedi}, \citenamefont {Raina},\
  and\ \citenamefont {Hirsch}}]{Rath2009}%
  \BibitemOpen
  \bibfield  {author} {\bibinfo {author} {\bibfnamefont {P.~K.}\ \bibnamefont
  {Rath}}, \bibinfo {author} {\bibfnamefont {R.}~\bibnamefont {Chandra}},
  \bibinfo {author} {\bibfnamefont {K.}~\bibnamefont {Chaturvedi}}, \bibinfo
  {author} {\bibfnamefont {P.~K.}\ \bibnamefont {Raina}}, \ and\ \bibinfo
  {author} {\bibfnamefont {J.~G.}\ \bibnamefont {Hirsch}},\ }\href {\doibase
  10.1103/physrevc.80.044303} {\bibfield  {journal} {\bibinfo  {journal} {Phys.
  Rev. C}\ }\textbf {\bibinfo {volume} {80}},\ \bibinfo {pages} {044303}
  (\bibinfo {year} {2009})}\BibitemShut {NoStop}%
\bibitem [{\citenamefont {Rath}\ \emph {et~al.}(2010)\citenamefont {Rath},
  \citenamefont {Chandra}, \citenamefont {Chaturvedi}, \citenamefont {Raina},\
  and\ \citenamefont {Hirsch}}]{Rath2010}%
  \BibitemOpen
  \bibfield  {author} {\bibinfo {author} {\bibfnamefont {P.~K.}\ \bibnamefont
  {Rath}}, \bibinfo {author} {\bibfnamefont {R.}~\bibnamefont {Chandra}},
  \bibinfo {author} {\bibfnamefont {K.}~\bibnamefont {Chaturvedi}}, \bibinfo
  {author} {\bibfnamefont {P.~K.}\ \bibnamefont {Raina}}, \ and\ \bibinfo
  {author} {\bibfnamefont {J.~G.}\ \bibnamefont {Hirsch}},\ }\href {\doibase
  10.1103/physrevc.82.064310} {\bibfield  {journal} {\bibinfo  {journal} {Phys.
  Rev. C}\ }\textbf {\bibinfo {volume} {82}},\ \bibinfo {pages} {064310}
  (\bibinfo {year} {2010})}\BibitemShut {NoStop}%
\bibitem [{\citenamefont {Rath}\ \emph {et~al.}(2012)\citenamefont {Rath},
  \citenamefont {Chandra}, \citenamefont {Raina}, \citenamefont {Chaturvedi},\
  and\ \citenamefont {Hirsch}}]{Rath2012}%
  \BibitemOpen
  \bibfield  {author} {\bibinfo {author} {\bibfnamefont {P.~K.}\ \bibnamefont
  {Rath}}, \bibinfo {author} {\bibfnamefont {R.}~\bibnamefont {Chandra}},
  \bibinfo {author} {\bibfnamefont {P.~K.}\ \bibnamefont {Raina}}, \bibinfo
  {author} {\bibfnamefont {K.}~\bibnamefont {Chaturvedi}}, \ and\ \bibinfo
  {author} {\bibfnamefont {J.~G.}\ \bibnamefont {Hirsch}},\ }\href {\doibase
  10.1103/physrevc.85.014308} {\bibfield  {journal} {\bibinfo  {journal} {Phys.
  Rev. C}\ }\textbf {\bibinfo {volume} {85}},\ \bibinfo {pages} {014308}
  (\bibinfo {year} {2012})}\BibitemShut {NoStop}%
\bibitem [{\citenamefont {Rath}\ \emph {et~al.}(2013)\citenamefont {Rath},
  \citenamefont {Chandra}, \citenamefont {Chaturvedi}, \citenamefont {Lohani},
  \citenamefont {Raina},\ and\ \citenamefont {Hirsch}}]{Rath2013}%
  \BibitemOpen
  \bibfield  {author} {\bibinfo {author} {\bibfnamefont {P.~K.}\ \bibnamefont
  {Rath}}, \bibinfo {author} {\bibfnamefont {R.}~\bibnamefont {Chandra}},
  \bibinfo {author} {\bibfnamefont {K.}~\bibnamefont {Chaturvedi}}, \bibinfo
  {author} {\bibfnamefont {P.}~\bibnamefont {Lohani}}, \bibinfo {author}
  {\bibfnamefont {P.~K.}\ \bibnamefont {Raina}}, \ and\ \bibinfo {author}
  {\bibfnamefont {J.~G.}\ \bibnamefont {Hirsch}},\ }\href {\doibase
  10.1103/physrevc.88.064322} {\bibfield  {journal} {\bibinfo  {journal} {Phys.
  Rev. C}\ }\textbf {\bibinfo {volume} {88}},\ \bibinfo {pages} {064322}
  (\bibinfo {year} {2013})}\BibitemShut {NoStop}%
\bibitem [{\citenamefont {Barea}\ and\ \citenamefont
  {Iachello}(2009)}]{Barea2009}%
  \BibitemOpen
  \bibfield  {author} {\bibinfo {author} {\bibfnamefont {J.}~\bibnamefont
  {Barea}}\ and\ \bibinfo {author} {\bibfnamefont {F.}~\bibnamefont
  {Iachello}},\ }\href {\doibase 10.1103/physrevc.79.044301} {\bibfield
  {journal} {\bibinfo  {journal} {Phys. Rev. C}\ }\textbf {\bibinfo {volume}
  {79}},\ \bibinfo {pages} {044301} (\bibinfo {year} {2009})}\BibitemShut
  {NoStop}%
\bibitem [{\citenamefont {Barea}\ \emph {et~al.}(2012)\citenamefont {Barea},
  \citenamefont {Kotila},\ and\ \citenamefont {Iachello}}]{Barea2012}%
  \BibitemOpen
  \bibfield  {author} {\bibinfo {author} {\bibfnamefont {J.}~\bibnamefont
  {Barea}}, \bibinfo {author} {\bibfnamefont {J.}~\bibnamefont {Kotila}}, \
  and\ \bibinfo {author} {\bibfnamefont {F.}~\bibnamefont {Iachello}},\ }\href
  {\doibase 10.1103/physrevlett.109.042501} {\bibfield  {journal} {\bibinfo
  {journal} {Phys. Rev. Lett.}\ }\textbf {\bibinfo {volume} {109}},\ \bibinfo
  {pages} {042501} (\bibinfo {year} {2012})}\BibitemShut {NoStop}%
\bibitem [{\citenamefont {Barea}\ \emph {et~al.}(2013)\citenamefont {Barea},
  \citenamefont {Kotila},\ and\ \citenamefont {Iachello}}]{Barea2013}%
  \BibitemOpen
  \bibfield  {author} {\bibinfo {author} {\bibfnamefont {J.}~\bibnamefont
  {Barea}}, \bibinfo {author} {\bibfnamefont {J.}~\bibnamefont {Kotila}}, \
  and\ \bibinfo {author} {\bibfnamefont {F.}~\bibnamefont {Iachello}},\ }\href
  {\doibase 10.1103/physrevc.87.014315} {\bibfield  {journal} {\bibinfo
  {journal} {Phys. Rev. C}\ }\textbf {\bibinfo {volume} {87}},\ \bibinfo
  {pages} {014315} (\bibinfo {year} {2013})}\BibitemShut {NoStop}%
\bibitem [{\citenamefont {Barea}\ \emph {et~al.}(2015)\citenamefont {Barea},
  \citenamefont {Kotila},\ and\ \citenamefont {Iachello}}]{Barea2015}%
  \BibitemOpen
  \bibfield  {author} {\bibinfo {author} {\bibfnamefont {J.}~\bibnamefont
  {Barea}}, \bibinfo {author} {\bibfnamefont {J.}~\bibnamefont {Kotila}}, \
  and\ \bibinfo {author} {\bibfnamefont {F.}~\bibnamefont {Iachello}},\ }\href
  {\doibase 10.1103/PhysRevC.91.034304} {\bibfield  {journal} {\bibinfo
  {journal} {Phys. Rev. C}\ }\textbf {\bibinfo {volume} {91}},\ \bibinfo
  {pages} {034304} (\bibinfo {year} {2015})}\BibitemShut {NoStop}%
\bibitem [{\citenamefont {Rodr{\'{\i}}guez}\ and\ \citenamefont
  {Mart{\'{\i}}nez-Pinedo}(2010)}]{Rodriguez2010}%
  \BibitemOpen
  \bibfield  {author} {\bibinfo {author} {\bibfnamefont {T.~R.}\ \bibnamefont
  {Rodr{\'{\i}}guez}}\ and\ \bibinfo {author} {\bibfnamefont {G.}~\bibnamefont
  {Mart{\'{\i}}nez-Pinedo}},\ }\href {\doibase 10.1103/physrevlett.105.252503}
  {\bibfield  {journal} {\bibinfo  {journal} {Phys. Rev. Lett.}\ }\textbf
  {\bibinfo {volume} {105}},\ \bibinfo {pages} {252503} (\bibinfo {year}
  {2010})}\BibitemShut {NoStop}%
\bibitem [{\citenamefont {Rodr{\'{\i}}guez}\ and\ \citenamefont
  {Mart{\'{\i}}nez-Pinedo}(2011)}]{Rodriguez2011}%
  \BibitemOpen
  \bibfield  {author} {\bibinfo {author} {\bibfnamefont {T.~R.}\ \bibnamefont
  {Rodr{\'{\i}}guez}}\ and\ \bibinfo {author} {\bibfnamefont {G.}~\bibnamefont
  {Mart{\'{\i}}nez-Pinedo}},\ }\href {\doibase 10.1016/j.ppnp.2011.01.047}
  {\bibfield  {journal} {\bibinfo  {journal} {Prog. Part. Nucl. Phys.}\
  }\textbf {\bibinfo {volume} {66}},\ \bibinfo {pages} {436} (\bibinfo {year}
  {2011})}\BibitemShut {NoStop}%
\bibitem [{\citenamefont {Vaquero}\ \emph {et~al.}(2013)\citenamefont
  {Vaquero}, \citenamefont {Rodr{\'{\i}}guez},\ and\ \citenamefont
  {Egido}}]{Vaquero2013}%
  \BibitemOpen
  \bibfield  {author} {\bibinfo {author} {\bibfnamefont {N.~L.}\ \bibnamefont
  {Vaquero}}, \bibinfo {author} {\bibfnamefont {T.~R.}\ \bibnamefont
  {Rodr{\'{\i}}guez}}, \ and\ \bibinfo {author} {\bibfnamefont {J.~L.}\
  \bibnamefont {Egido}},\ }\href {\doibase 10.1103/physrevlett.111.142501}
  {\bibfield  {journal} {\bibinfo  {journal} {Phys. Rev. Lett.}\ }\textbf
  {\bibinfo {volume} {111}},\ \bibinfo {pages} {142501} (\bibinfo {year}
  {2013})}\BibitemShut {NoStop}%
\bibitem [{\citenamefont {{\v{S}}imkovic}\ \emph
  {et~al.}(2009{\natexlab{b}})\citenamefont {{\v{S}}imkovic}, \citenamefont
  {Faessler}, \citenamefont {M{\"u}ther}, \citenamefont {Rodin},\ and\
  \citenamefont {Stauf}}]{Simkovic2009a}%
  \BibitemOpen
  \bibfield  {author} {\bibinfo {author} {\bibfnamefont {F.}~\bibnamefont
  {{\v{S}}imkovic}}, \bibinfo {author} {\bibfnamefont {A.}~\bibnamefont
  {Faessler}}, \bibinfo {author} {\bibfnamefont {H.}~\bibnamefont
  {M{\"u}ther}}, \bibinfo {author} {\bibfnamefont {V.}~\bibnamefont {Rodin}}, \
  and\ \bibinfo {author} {\bibfnamefont {M.}~\bibnamefont {Stauf}},\ }\href
  {\doibase 10.1103/physrevc.79.055501} {\bibfield  {journal} {\bibinfo
  {journal} {Phys. Rev. C}\ }\textbf {\bibinfo {volume} {79}},\ \bibinfo
  {pages} {055501} (\bibinfo {year} {2009}{\natexlab{b}})}\BibitemShut
  {NoStop}%
\bibitem [{\citenamefont {Jastrow}(1955)}]{Jastrow1955}%
  \BibitemOpen
  \bibfield  {author} {\bibinfo {author} {\bibfnamefont {R.}~\bibnamefont
  {Jastrow}},\ }\href@noop {} {\bibfield  {journal} {\bibinfo  {journal} {Phys.
  Rev.}\ }\textbf {\bibinfo {volume} {98}},\ \bibinfo {pages} {1479} (\bibinfo
  {year} {1955})}\BibitemShut {NoStop}%
\bibitem [{\citenamefont {Miller}\ and\ \citenamefont
  {Spencer}(1976)}]{Miller1976}%
  \BibitemOpen
  \bibfield  {author} {\bibinfo {author} {\bibfnamefont {G.~A.}\ \bibnamefont
  {Miller}}\ and\ \bibinfo {author} {\bibfnamefont {J.~E.}\ \bibnamefont
  {Spencer}},\ }\href {\doibase http://dx.doi.org/10.1016/0003-4916(76)90073-7}
  {\bibfield  {journal} {\bibinfo  {journal} {Ann. Phys.}\ }\textbf {\bibinfo
  {volume} {100}},\ \bibinfo {pages} {562} (\bibinfo {year}
  {1976})}\BibitemShut {NoStop}%
\bibitem [{\citenamefont {Yao}\ \emph {et~al.}(2008)\citenamefont {Yao},
  \citenamefont {Meng}, \citenamefont {Pe\~na Arteaga},\ and\ \citenamefont
  {Ring}}]{Yao2008}%
  \BibitemOpen
  \bibfield  {author} {\bibinfo {author} {\bibfnamefont {J.~M.}\ \bibnamefont
  {Yao}}, \bibinfo {author} {\bibfnamefont {J.}~\bibnamefont {Meng}}, \bibinfo
  {author} {\bibfnamefont {D.}\ \bibnamefont {Pe\~na Arteaga}}, \ and\ \bibinfo
  {author} {\bibfnamefont {P.}~\bibnamefont {Ring}},\ }\href
  {http://stacks.iop.org/0256-307X/25/i=10/a=024} {\bibfield  {journal}
  {\bibinfo  {journal} {Chin. Phys. Lett.}\ }\textbf {\bibinfo {volume} {25}},\
  \bibinfo {pages} {3609} (\bibinfo {year} {2008})}\BibitemShut {NoStop}%
\bibitem [{\citenamefont {Yao}\ \emph {et~al.}(2009)\citenamefont {Yao},
  \citenamefont {Meng}, \citenamefont {Ring},\ and\ \citenamefont
  {Pe\~na Arteaga}}]{Yao2009}%
  \BibitemOpen
  \bibfield  {author} {\bibinfo {author} {\bibfnamefont {J.~M.}\ \bibnamefont
  {Yao}}, \bibinfo {author} {\bibfnamefont {J.}~\bibnamefont {Meng}}, \bibinfo
  {author} {\bibfnamefont {P.}~\bibnamefont {Ring}}, \ and\ \bibinfo {author}
  {\bibfnamefont {D.}\ \bibnamefont {Pe\~na Arteaga}},\ }\href {\doibase
  10.1103/physrevc.79.044312} {\bibfield  {journal} {\bibinfo  {journal} {Phys.
  Rev. C}\ }\textbf {\bibinfo {volume} {79}},\ \bibinfo {pages} {044312}
  (\bibinfo {year} {2009})}\BibitemShut {NoStop}%
\bibitem [{\citenamefont {Yao}\ \emph {et~al.}(2010)\citenamefont {Yao},
  \citenamefont {Meng}, \citenamefont {Ring},\ and\ \citenamefont
  {Vretenar}}]{Yao2010}%
  \BibitemOpen
  \bibfield  {author} {\bibinfo {author} {\bibfnamefont {J.~M.}\ \bibnamefont
  {Yao}}, \bibinfo {author} {\bibfnamefont {J.}~\bibnamefont {Meng}}, \bibinfo
  {author} {\bibfnamefont {P.}~\bibnamefont {Ring}}, \ and\ \bibinfo {author}
  {\bibfnamefont {D.}~\bibnamefont {Vretenar}},\ }\href {\doibase
  10.1103/physrevc.81.044311} {\bibfield  {journal} {\bibinfo  {journal} {Phys.
  Rev. C}\ }\textbf {\bibinfo {volume} {81}},\ \bibinfo {pages} {044311}
  (\bibinfo {year} {2010})}\BibitemShut {NoStop}%
\bibitem [{\citenamefont {Yao}\ \emph {et~al.}(2014)\citenamefont {Yao},
  \citenamefont {Hagino}, \citenamefont {Li}, \citenamefont {Meng},\ and\
  \citenamefont {Ring}}]{Yao2014}%
  \BibitemOpen
  \bibfield  {author} {\bibinfo {author} {\bibfnamefont {J.~M.}\ \bibnamefont
  {Yao}}, \bibinfo {author} {\bibfnamefont {K.}~\bibnamefont {Hagino}},
  \bibinfo {author} {\bibfnamefont {Z.~P.}\ \bibnamefont {Li}}, \bibinfo
  {author} {\bibfnamefont {J.}~\bibnamefont {Meng}}, \ and\ \bibinfo {author}
  {\bibfnamefont {P.}~\bibnamefont {Ring}},\ }\href {\doibase
  10.1103/physrevc.89.054306} {\bibfield  {journal} {\bibinfo  {journal} {Phys.
  Rev. C}\ }\textbf {\bibinfo {volume} {89}},\ \bibinfo {pages} {054306}
  (\bibinfo {year} {2014})}\BibitemShut {NoStop}%
\bibitem [{\citenamefont {Yao}\ \emph {et~al.}(2015{\natexlab{b}})\citenamefont
  {Yao}, \citenamefont {Zhou},\ and\ \citenamefont {Li}}]{Yao2015a}%
  \BibitemOpen
  \bibfield  {author} {\bibinfo {author} {\bibfnamefont {J.~M.}\ \bibnamefont
  {Yao}}, \bibinfo {author} {\bibfnamefont {E.~F.}\ \bibnamefont {Zhou}}, \
  and\ \bibinfo {author} {\bibfnamefont {Z.~P.}\ \bibnamefont {Li}},\ }\href
  {\doibase 10.1103/PhysRevC.92.041304} {\bibfield  {journal} {\bibinfo
  {journal} {Phys. Rev. C}\ }\textbf {\bibinfo {volume} {92}},\ \bibinfo
  {pages} {041304} (\bibinfo {year} {2015}{\natexlab{b}})}\BibitemShut
  {NoStop}%
\bibitem [{\citenamefont {Zhao}\ \emph {et~al.}(2010)\citenamefont {Zhao},
  \citenamefont {Li}, \citenamefont {Yao},\ and\ \citenamefont
  {Meng}}]{Zhao2010}%
  \BibitemOpen
  \bibfield  {author} {\bibinfo {author} {\bibfnamefont {P.~W.}\ \bibnamefont
  {Zhao}}, \bibinfo {author} {\bibfnamefont {Z.~P.}\ \bibnamefont {Li}},
  \bibinfo {author} {\bibfnamefont {J.~M.}\ \bibnamefont {Yao}}, \ and\
  \bibinfo {author} {\bibfnamefont {J.}~\bibnamefont {Meng}},\ }\href {\doibase
  10.1103/physrevc.82.054319} {\bibfield  {journal} {\bibinfo  {journal} {Phys.
  Rev. C}\ }\textbf {\bibinfo {volume} {82}},\ \bibinfo {pages} {054319}
  (\bibinfo {year} {2010})}\BibitemShut {NoStop}%
\bibitem [{\citenamefont {Ring}\ and\ \citenamefont {Schuck}(1980)}]{Ring1980}%
  \BibitemOpen
  \bibfield  {author} {\bibinfo {author} {\bibfnamefont {P.}~\bibnamefont
  {Ring}}\ and\ \bibinfo {author} {\bibfnamefont {P.}~\bibnamefont {Schuck}},\
  }\href@noop {} {\emph {\bibinfo {title} {The Nuclear Many-Body Problem}}}\
  (\bibinfo  {publisher} {Springer-Verlag New York Inc.},\ \bibinfo {year}
  {1980})\BibitemShut {NoStop}%
\bibitem [{\citenamefont {Gambhir}\ \emph {et~al.}(1990)\citenamefont
  {Gambhir}, \citenamefont {Ring},\ and\ \citenamefont {Thimet}}]{Gambhir1990}%
  \BibitemOpen
  \bibfield  {author} {\bibinfo {author} {\bibfnamefont {Y.}~\bibnamefont
  {Gambhir}}, \bibinfo {author} {\bibfnamefont {P.}~\bibnamefont {Ring}}, \
  and\ \bibinfo {author} {\bibfnamefont {A.}~\bibnamefont {Thimet}},\ }\href
  {\doibase http://dx.doi.org/10.1016/0003-4916(90)90330-Q} {\bibfield
  {journal} {\bibinfo  {journal} {Ann. Phys.}\ }\textbf {\bibinfo {volume}
  {198}},\ \bibinfo {pages} {132} (\bibinfo {year} {1990})}\BibitemShut
  {NoStop}%
\bibitem [{\citenamefont {Tian}\ \emph {et~al.}(2009)\citenamefont {Tian},
  \citenamefont {Ma},\ and\ \citenamefont {Ring}}]{Tian2009}%
  \BibitemOpen
  \bibfield  {author} {\bibinfo {author} {\bibfnamefont {Y.}~\bibnamefont
  {Tian}}, \bibinfo {author} {\bibfnamefont {Z.}~\bibnamefont {Ma}}, \ and\
  \bibinfo {author} {\bibfnamefont {P.}~\bibnamefont {Ring}},\ }\href {\doibase
  http://dx.doi.org/10.1016/j.physletb.2009.04.067} {\bibfield  {journal}
  {\bibinfo  {journal} {Phys. Lett. B}\ }\textbf {\bibinfo {volume} {676}},\
  \bibinfo {pages} {44} (\bibinfo {year} {2009})}\BibitemShut {NoStop}%
\bibitem [{\citenamefont {Suhonen}(1993)}]{Suhonen1993}%
  \BibitemOpen
  \bibfield  {author} {\bibinfo {author} {\bibfnamefont {J.}~\bibnamefont
  {Suhonen}},\ }\href {http://stacks.iop.org/0954-3899/19/i=1/a=010} {\bibfield
   {journal} {\bibinfo  {journal} {Journal of Physics G: Nuclear and Particle
  Physics}\ }\textbf {\bibinfo {volume} {19}},\ \bibinfo {pages} {139}
  (\bibinfo {year} {1993})}\BibitemShut {NoStop}%
\bibitem [{\citenamefont {Rodin}\ \emph {et~al.}(2003)\citenamefont {Rodin},
  \citenamefont {Faessler}, \citenamefont {{\v{S}}imkovic},\ and\ \citenamefont
  {Vogel}}]{Rodin2003}%
  \BibitemOpen
  \bibfield  {author} {\bibinfo {author} {\bibfnamefont {V.~A.}\ \bibnamefont
  {Rodin}}, \bibinfo {author} {\bibfnamefont {A.}~\bibnamefont {Faessler}},
  \bibinfo {author} {\bibfnamefont {F.}~\bibnamefont {{\v{S}}imkovic}}, \ and\
  \bibinfo {author} {\bibfnamefont {P.}~\bibnamefont {Vogel}},\ }\href
  {\doibase 10.1103/physrevc.68.044302} {\bibfield  {journal} {\bibinfo
  {journal} {Phys. Rev. C}\ }\textbf {\bibinfo {volume} {68}},\ \bibinfo
  {pages} {044302} (\bibinfo {year} {2003})}\BibitemShut {NoStop}%
\bibitem [{\citenamefont {Hinohara}\ and\ \citenamefont
  {Engel}(2014)}]{Hinohara2014}%
  \BibitemOpen
  \bibfield  {author} {\bibinfo {author} {\bibfnamefont {N.}~\bibnamefont
  {Hinohara}}\ and\ \bibinfo {author} {\bibfnamefont {J.}~\bibnamefont
  {Engel}},\ }\href {\doibase 10.1103/physrevc.90.031301} {\bibfield  {journal}
  {\bibinfo  {journal} {Phys. Rev. C}\ }\textbf {\bibinfo {volume} {90}},\
  \bibinfo {pages} {031301} (\bibinfo {year} {2014})}\BibitemShut {NoStop}%
\bibitem [{\citenamefont {Men\'endez}\ \emph {et~al.}(2016)\citenamefont
  {Men\'endez}, \citenamefont {Hinohara}, \citenamefont {Engel}, \citenamefont
  {Mart\'{\i}nez-Pinedo},\ and\ \citenamefont {Rodr\'{\i}guez}}]{Menendez2016}%
  \BibitemOpen
  \bibfield  {author} {\bibinfo {author} {\bibfnamefont {J.}~\bibnamefont
  {Men\'endez}}, \bibinfo {author} {\bibfnamefont {N.}~\bibnamefont
  {Hinohara}}, \bibinfo {author} {\bibfnamefont {J.}~\bibnamefont {Engel}},
  \bibinfo {author} {\bibfnamefont {G.}~\bibnamefont {Mart\'{\i}nez-Pinedo}}, \
  and\ \bibinfo {author} {\bibfnamefont {T.~R.}\ \bibnamefont
  {Rodr\'{\i}guez}},\ }\href {\doibase 10.1103/PhysRevC.93.014305} {\bibfield
  {journal} {\bibinfo  {journal} {Phys. Rev. C}\ }\textbf {\bibinfo {volume}
  {93}},\ \bibinfo {pages} {014305} (\bibinfo {year} {2016})}\BibitemShut
  {NoStop}%
\bibitem [{\citenamefont {Ericson}\ and\ \citenamefont
  {Weise}(1988)}]{Ericson1988}%
  \BibitemOpen
  \bibfield  {author} {\bibinfo {author} {\bibfnamefont {T.}~\bibnamefont
  {Ericson}}\ and\ \bibinfo {author} {\bibfnamefont {W.}~\bibnamefont
  {Weise}},\ }\href {https://books.google.com/books?id=v099AAAAIAAJ} {\emph
  {\bibinfo {title} {Pions and nuclei}}},\ Oxford Science Publications\
  (\bibinfo  {publisher} {Clarendon Press},\ \bibinfo {year}
  {1988})\BibitemShut {NoStop}%
\bibitem [{\citenamefont {Kitagaki}\ \emph {et~al.}(1983)\citenamefont
  {Kitagaki}, \citenamefont {Tanaka}, \citenamefont {Yuta}, \citenamefont
  {Abe}, \citenamefont {Hasegawa}, \citenamefont {Yamaguchi}, \citenamefont
  {Tamai}, \citenamefont {Hayashino}, \citenamefont {Otani}, \citenamefont
  {Hayano}, \citenamefont {Sagawa}, \citenamefont {Burnstein}, \citenamefont
  {Hanlon}, \citenamefont {Rubin}, \citenamefont {Chang}, \citenamefont
  {Kunori}, \citenamefont {Snow}, \citenamefont {Son}, \citenamefont
  {Steinberg}, \citenamefont {Zieminska}, \citenamefont {Engelmann},
  \citenamefont {Kafka}, \citenamefont {Sommars}, \citenamefont {Chang},
  \citenamefont {Mann}, \citenamefont {Napier},\ and\ \citenamefont
  {Schneps}}]{Kitagaki1983}%
  \BibitemOpen
  \bibfield  {author} {\bibinfo {author} {\bibfnamefont {T.}~\bibnamefont
  {Kitagaki}}, \bibinfo {author} {\bibfnamefont {S.}~\bibnamefont {Tanaka}},
  \bibinfo {author} {\bibfnamefont {H.}~\bibnamefont {Yuta}}, \bibinfo {author}
  {\bibfnamefont {K.}~\bibnamefont {Abe}}, \bibinfo {author} {\bibfnamefont
  {K.}~\bibnamefont {Hasegawa}}, \bibinfo {author} {\bibfnamefont
  {A.}~\bibnamefont {Yamaguchi}}, \bibinfo {author} {\bibfnamefont
  {K.}~\bibnamefont {Tamai}}, \bibinfo {author} {\bibfnamefont
  {T.}~\bibnamefont {Hayashino}}, \bibinfo {author} {\bibfnamefont
  {Y.}~\bibnamefont {Otani}}, \bibinfo {author} {\bibfnamefont
  {H.}~\bibnamefont {Hayano}}, \bibinfo {author} {\bibfnamefont
  {H.}~\bibnamefont {Sagawa}}, \bibinfo {author} {\bibfnamefont {R.~A.}\
  \bibnamefont {Burnstein}}, \bibinfo {author} {\bibfnamefont {J.}~\bibnamefont
  {Hanlon}}, \bibinfo {author} {\bibfnamefont {H.~A.}\ \bibnamefont {Rubin}},
  \bibinfo {author} {\bibfnamefont {C.~Y.}\ \bibnamefont {Chang}}, \bibinfo
  {author} {\bibfnamefont {S.}~\bibnamefont {Kunori}}, \bibinfo {author}
  {\bibfnamefont {G.~A.}\ \bibnamefont {Snow}}, \bibinfo {author}
  {\bibfnamefont {D.}~\bibnamefont {Son}}, \bibinfo {author} {\bibfnamefont
  {P.~H.}\ \bibnamefont {Steinberg}}, \bibinfo {author} {\bibfnamefont
  {D.}~\bibnamefont {Zieminska}}, \bibinfo {author} {\bibfnamefont
  {R.}~\bibnamefont {Engelmann}}, \bibinfo {author} {\bibfnamefont
  {T.}~\bibnamefont {Kafka}}, \bibinfo {author} {\bibfnamefont
  {S.}~\bibnamefont {Sommars}}, \bibinfo {author} {\bibfnamefont {C.~C.}\
  \bibnamefont {Chang}}, \bibinfo {author} {\bibfnamefont {W.~A.}\ \bibnamefont
  {Mann}}, \bibinfo {author} {\bibfnamefont {A.}~\bibnamefont {Napier}}, \ and\
  \bibinfo {author} {\bibfnamefont {J.}~\bibnamefont {Schneps}},\ }\href
  {\doibase 10.1103/PhysRevD.28.436} {\bibfield  {journal} {\bibinfo  {journal}
  {Phys. Rev. D}\ }\textbf {\bibinfo {volume} {28}},\ \bibinfo {pages} {436}
  (\bibinfo {year} {1983})}\BibitemShut {NoStop}%
\bibitem [{\citenamefont {{\v{S}}imkovic}\ \emph {et~al.}(1992)\citenamefont
  {{\v{S}}imkovic}, \citenamefont {Efimov}, \citenamefont {Ivanov},\ and\
  \citenamefont {Lyubovitskij}}]{Simkovic1992}%
  \BibitemOpen
  \bibfield  {author} {\bibinfo {author} {\bibfnamefont {F.}~\bibnamefont
  {{\v{S}}imkovic}}, \bibinfo {author} {\bibfnamefont {G.~V.}\ \bibnamefont
  {Efimov}}, \bibinfo {author} {\bibfnamefont {M.~A.}\ \bibnamefont {Ivanov}},
  \ and\ \bibinfo {author} {\bibfnamefont {V.~E.}\ \bibnamefont
  {Lyubovitskij}},\ }\href {\doibase 10.1007/BF01298480} {\bibfield  {journal}
  {\bibinfo  {journal} {Z. Phys. A}\
  }\textbf {\bibinfo {volume} {341}},\ \bibinfo {pages} {193} (\bibinfo {year}
  {1992})}\BibitemShut {NoStop}%
\bibitem [{\citenamefont {Vergados}(1982)}]{Vergados1982}%
  \BibitemOpen
  \bibfield  {author} {\bibinfo {author} {\bibfnamefont {J.~D.}\ \bibnamefont
  {Vergados}},\ }\href {\doibase 10.1103/PhysRevD.25.914} {\bibfield  {journal}
  {\bibinfo  {journal} {Phys. Rev. D}\ }\textbf {\bibinfo {volume} {25}},\
  \bibinfo {pages} {914} (\bibinfo {year} {1982})}\BibitemShut {NoStop}%
\bibitem [{\citenamefont {Faessler}\ \emph {et~al.}(1997)\citenamefont
  {Faessler}, \citenamefont {Kovalenko}, \citenamefont {\ifmmode~\check{S}\else
  \v{S}\fi{}imkovic},\ and\ \citenamefont {Schwieger}}]{Faessler1997}%
  \BibitemOpen
  \bibfield  {author} {\bibinfo {author} {\bibfnamefont {A.}~\bibnamefont
  {Faessler}}, \bibinfo {author} {\bibfnamefont {S.}~\bibnamefont {Kovalenko}},
  \bibinfo {author} {\bibfnamefont {F.}~\bibnamefont {\ifmmode~\check{S}\else
  \v{S}\fi{}imkovic}}, \ and\ \bibinfo {author} {\bibfnamefont
  {J.}~\bibnamefont {Schwieger}},\ }\href {\doibase 10.1103/PhysRevLett.78.183}
  {\bibfield  {journal} {\bibinfo  {journal} {Phys. Rev. Lett.}\ }\textbf
  {\bibinfo {volume} {78}},\ \bibinfo {pages} {183} (\bibinfo {year}
  {1997})}\BibitemShut {NoStop}%
\bibitem [{\citenamefont {Pr\'ezeau}\ \emph {et~al.}(2003)\citenamefont
  {Pr\'ezeau}, \citenamefont {Ramsey-Musolf},\ and\ \citenamefont
  {Vogel}}]{Prezeau2003}%
  \BibitemOpen
  \bibfield  {author} {\bibinfo {author} {\bibfnamefont {G.}~\bibnamefont
  {Pr\'ezeau}}, \bibinfo {author} {\bibfnamefont {M.}~\bibnamefont
  {Ramsey-Musolf}}, \ and\ \bibinfo {author} {\bibfnamefont {P.}~\bibnamefont
  {Vogel}},\ }\href {\doibase 10.1103/PhysRevD.68.034016} {\bibfield  {journal}
  {\bibinfo  {journal} {Phys. Rev. D}\ }\textbf {\bibinfo {volume} {68}},\
  \bibinfo {pages} {034016} (\bibinfo {year} {2003})}\BibitemShut {NoStop}%
\bibitem [{\citenamefont {Umehara}\ \emph {et~al.}(2008)\citenamefont
  {Umehara}, \citenamefont {Kishimoto}, \citenamefont {Ogawa}, \citenamefont
  {Hazama}, \citenamefont {Miyawaki}, \citenamefont {Yoshida}, \citenamefont
  {Matsuoka}, \citenamefont {Kishimoto}, \citenamefont {Katsuki}, \citenamefont
  {Sakai}, \citenamefont {Yokoyama}, \citenamefont {Mukaida}, \citenamefont
  {Tomii}, \citenamefont {Tatewaki}, \citenamefont {Kobayashi},\ and\
  \citenamefont {Yanagisawa}}]{Umehara2008}%
  \BibitemOpen
  \bibfield  {author} {\bibinfo {author} {\bibfnamefont {S.}~\bibnamefont
  {Umehara}}, \bibinfo {author} {\bibfnamefont {T.}~\bibnamefont {Kishimoto}},
  \bibinfo {author} {\bibfnamefont {I.}~\bibnamefont {Ogawa}}, \bibinfo
  {author} {\bibfnamefont {R.}~\bibnamefont {Hazama}}, \bibinfo {author}
  {\bibfnamefont {H.}~\bibnamefont {Miyawaki}}, \bibinfo {author}
  {\bibfnamefont {S.}~\bibnamefont {Yoshida}}, \bibinfo {author} {\bibfnamefont
  {K.}~\bibnamefont {Matsuoka}}, \bibinfo {author} {\bibfnamefont
  {K.}~\bibnamefont {Kishimoto}}, \bibinfo {author} {\bibfnamefont
  {A.}~\bibnamefont {Katsuki}}, \bibinfo {author} {\bibfnamefont
  {H.}~\bibnamefont {Sakai}}, \bibinfo {author} {\bibfnamefont
  {D.}~\bibnamefont {Yokoyama}}, \bibinfo {author} {\bibfnamefont
  {K.}~\bibnamefont {Mukaida}}, \bibinfo {author} {\bibfnamefont
  {S.}~\bibnamefont {Tomii}}, \bibinfo {author} {\bibfnamefont
  {Y.}~\bibnamefont {Tatewaki}}, \bibinfo {author} {\bibfnamefont
  {T.}~\bibnamefont {Kobayashi}}, \ and\ \bibinfo {author} {\bibfnamefont
  {A.}~\bibnamefont {Yanagisawa}},\ }\href {\doibase
  10.1103/physrevc.78.058501} {\bibfield  {journal} {\bibinfo  {journal} {Phys.
  Rev. C}\ }\textbf {\bibinfo {volume} {78}},\ \bibinfo {pages} {058501}
  (\bibinfo {year} {2008})}\BibitemShut {NoStop}%
\bibitem [{\citenamefont {Agostini{, \emph{et al.}}}(2013)}]{Agostini2013}%
  \BibitemOpen
  \bibfield  {author} {\bibinfo {author} {\bibfnamefont {M.}~\bibnamefont
  {Agostini{, \emph{et al.}}}} (\bibinfo {collaboration} {GERDA
  Collaboration}),\ }\href {\doibase 10.1103/physrevlett.111.122503} {\bibfield
   {journal} {\bibinfo  {journal} {Phys. Rev. Lett.}\ }\textbf {\bibinfo
  {volume} {111}},\ \bibinfo {pages} {122503} (\bibinfo {year}
  {2013})}\BibitemShut {NoStop}%
\bibitem [{\citenamefont {Barabash}\ \emph {et~al.}(2011)\citenamefont
  {Barabash}, \citenamefont {Brudanin},\ and\ \citenamefont {{NEMO
  Collaboration}}}]{Barabash2011b}%
  \BibitemOpen
  \bibfield  {author} {\bibinfo {author} {\bibfnamefont {A.~S.}\ \bibnamefont
  {Barabash}}, \bibinfo {author} {\bibfnamefont {V.~B.}\ \bibnamefont
  {Brudanin}}, \ and\ \bibinfo {author} {\bibnamefont {{NEMO Collaboration}}},\
  }\href@noop {} {\bibfield  {journal} {\bibinfo  {journal} {Phys. Atom.
  Nucl.}\ }\textbf {\bibinfo {volume} {74}},\ \bibinfo {pages} {312} (\bibinfo
  {year} {2011})}\BibitemShut {NoStop}%
\bibitem [{\citenamefont {Argyriades{, \emph{et al.}}}(2010)}]{Argyriades2010}%
  \BibitemOpen
  \bibfield  {author} {\bibinfo {author} {\bibfnamefont {J.}~\bibnamefont
  {Argyriades{, \emph{et al.}}}} (\bibinfo {collaboration} {NEMO-3
  Collaboration}),\ }\href {\doibase
  http://dx.doi.org/10.1016/j.nuclphysa.2010.07.009} {\bibfield  {journal}
  {\bibinfo  {journal} {Nucl. Phys. A}\ }\textbf {\bibinfo {volume} {847}},\
  \bibinfo {pages} {168} (\bibinfo {year} {2010})}\BibitemShut {NoStop}%
\bibitem [{\citenamefont {Arnold{, \emph{et al.}}}(2014)}]{Arnold2014}%
  \BibitemOpen
  \bibfield  {author} {\bibinfo {author} {\bibfnamefont {R.}~\bibnamefont
  {Arnold{, \emph{et al.}}}} (\bibinfo {collaboration} {NEMO-3
  Collaboration}),\ }\href {\doibase 10.1103/PhysRevD.89.111101} {\bibfield
  {journal} {\bibinfo  {journal} {Phys. Rev. D}\ }\textbf {\bibinfo {volume}
  {89}},\ \bibinfo {pages} {111101} (\bibinfo {year} {2014})}\BibitemShut
  {NoStop}%
\bibitem [{\citenamefont {Danevich}\ \emph {et~al.}(2003)\citenamefont
  {Danevich}, \citenamefont {Georgadze}, \citenamefont {Kobychev},
  \citenamefont {Kropivyansky}, \citenamefont {Nikolaiko}, \citenamefont
  {Ponkratenko}, \citenamefont {Tretyak}, \citenamefont {Zdesenko},
  \citenamefont {Zdesenko}, \citenamefont {Bizzeti}, \citenamefont {Fazzini},\
  and\ \citenamefont {Maurenzig}}]{Danevich2003}%
  \BibitemOpen
  \bibfield  {author} {\bibinfo {author} {\bibfnamefont {F.~A.}\ \bibnamefont
  {Danevich}}, \bibinfo {author} {\bibfnamefont {A.~S.}\ \bibnamefont
  {Georgadze}}, \bibinfo {author} {\bibfnamefont {V.~V.}\ \bibnamefont
  {Kobychev}}, \bibinfo {author} {\bibfnamefont {B.~N.}\ \bibnamefont
  {Kropivyansky}}, \bibinfo {author} {\bibfnamefont {A.~S.}\ \bibnamefont
  {Nikolaiko}}, \bibinfo {author} {\bibfnamefont {O.~A.}\ \bibnamefont
  {Ponkratenko}}, \bibinfo {author} {\bibfnamefont {V.~I.}\ \bibnamefont
  {Tretyak}}, \bibinfo {author} {\bibfnamefont {S.~Y.}\ \bibnamefont
  {Zdesenko}}, \bibinfo {author} {\bibfnamefont {Y.~G.}\ \bibnamefont
  {Zdesenko}}, \bibinfo {author} {\bibfnamefont {P.~G.}\ \bibnamefont
  {Bizzeti}}, \bibinfo {author} {\bibfnamefont {T.~F.}\ \bibnamefont
  {Fazzini}}, \ and\ \bibinfo {author} {\bibfnamefont {P.~R.}\ \bibnamefont
  {Maurenzig}},\ }\href {\doibase 10.1103/PhysRevC.68.035501} {\bibfield
  {journal} {\bibinfo  {journal} {Phys. Rev. C}\ }\textbf {\bibinfo {volume}
  {68}},\ \bibinfo {pages} {035501} (\bibinfo {year} {2003})}\BibitemShut
  {NoStop}%
\bibitem [{\citenamefont {Hwang}\ \emph {et~al.}(2009)\citenamefont {Hwang},
  \citenamefont {Kwon}, \citenamefont {Kim}, \citenamefont {Kwak},
  \citenamefont {Kim}, \citenamefont {Kim}, \citenamefont {Kim}, \citenamefont
  {Kim}, \citenamefont {Lee}, \citenamefont {Lee},\ and\ \citenamefont {{KIMS
  Collaboration}}}]{Hwang2009}%
  \BibitemOpen
  \bibfield  {author} {\bibinfo {author} {\bibfnamefont {M.}~\bibnamefont
  {Hwang}}, \bibinfo {author} {\bibfnamefont {Y.}~\bibnamefont {Kwon}},
  \bibinfo {author} {\bibfnamefont {H.}~\bibnamefont {Kim}}, \bibinfo {author}
  {\bibfnamefont {J.}~\bibnamefont {Kwak}}, \bibinfo {author} {\bibfnamefont
  {S.}~\bibnamefont {Kim}}, \bibinfo {author} {\bibfnamefont {S.}~\bibnamefont
  {Kim}}, \bibinfo {author} {\bibfnamefont {T.}~\bibnamefont {Kim}}, \bibinfo
  {author} {\bibfnamefont {S.}~\bibnamefont {Kim}}, \bibinfo {author}
  {\bibfnamefont {H.}~\bibnamefont {Lee}}, \bibinfo {author} {\bibfnamefont
  {M.}~\bibnamefont {Lee}}, \ and\ \bibinfo {author} {\bibnamefont {{KIMS
  Collaboration}}},\ }\href@noop {} {\bibfield  {journal} {\bibinfo  {journal}
  {Astropart. Phys.}\ }\textbf {\bibinfo {volume} {31}},\ \bibinfo {pages}
  {412} (\bibinfo {year} {2009})}\BibitemShut {NoStop}%
\bibitem [{\citenamefont {Barabash}(2011)}]{Barabash2011}%
  \BibitemOpen
  \bibfield  {author} {\bibinfo {author} {\bibfnamefont {A.~S.}\ \bibnamefont
  {Barabash}},\ }\href {\doibase 10.1134/s1063778811030070} {\bibfield
  {journal} {\bibinfo  {journal} {Phys. Atom. Nucl.}\ }\textbf {\bibinfo
  {volume} {74}},\ \bibinfo {pages} {603} (\bibinfo {year} {2011})}\BibitemShut
  {NoStop}%
\bibitem [{\citenamefont {Andreotti{, \emph{et al.}}}(2011)}]{Andreotti2011}%
  \BibitemOpen
  \bibfield  {author} {\bibinfo {author} {\bibfnamefont {E.}~\bibnamefont
  {Andreotti{, \emph{et al.}}}} (\bibinfo {collaboration} {CUORICINO
  Collaboration}),\ }\href {\doibase
  http://dx.doi.org/10.1016/j.astropartphys.2011.02.002} {\bibfield  {journal}
  {\bibinfo  {journal} {Astropart. Phys.}\ }\textbf {\bibinfo {volume} {34}},\
  \bibinfo {pages} {822} (\bibinfo {year} {2011})}\BibitemShut {NoStop}%
\end{thebibliography}
%

\end{CJK}
\end{document}